\documentclass[sn-mathphys]{sn-jnl}

\jyear{2023}%

\theoremstyle{thmstyleone}%
\theoremstyle{thmstyletwo}%

\theoremstyle{thmstylethree}%

\begin{document}

\title[Characterization of quantum benchmarks for quantum circuit mapping]{Interaction graph-based characterization of quantum benchmarks for improving quantum circuit mapping techniques}


\author*[1,2]{\fnm{Medina} \sur{Bandic}}\email{m.bandic@tudelft.nl}

\author[3]{\fnm{Carmen } 
\sur{G. Almudever}}

\author[1,2]{\fnm{Sebastian} \sur{Feld}}

\affil*[1]{\orgname{Delft University of Technology}, \orgaddress{\country{The Netherlands}}}

\affil[2]{\orgname{QuTech}, \orgaddress{\city{Delft}, \country{The Netherlands}}}

\affil[3]{\orgname{Technical University of Valencia}, \orgaddress{\country{Spain}}}

\abstract{

To execute  quantum circuits on a quantum processor, they must be modified to meet the physical constraints of the quantum device.  This process, called \textit{quantum circuit mapping}, results in a gate/circuit depth overhead that depends on both the circuit properties and the hardware constraints, being the limited qubit connectivity a crucial restriction. In
this paper, we propose to extend the characterization of quantum circuits
by including qubit interaction graph properties using graph theory-based metrics in addition to previously used circuit-describing parameters. This approach allows for an in-depth analysis and clustering of quantum circuits and a comparison of performance when run on different quantum processors, aiding in developing better mapping techniques. Our study reveals a correlation between interaction graph-based parameters and mapping performance metrics for various existing configurations of quantum devices. We also provide a comprehensive collection of quantum circuits and algorithms for benchmarking future compilation techniques and quantum devices.

}


\maketitle

\section{Introduction}\label{sec1}

\label{Sec1}

Quantum technology has experienced rapid development in the past decades and has the potential to solve some classically intractable problems. Its contributions are still in the early stage, as current so-called Noisy Intermediate-Scale Quantum (NISQ) devices can only handle simple, small-sized algorithms considering they are limited by size and noise. They also encompass additional hardware constraints such as low qubit connectivity, reduced supported gate set, and limitations related to classical-control resources, which makes it even more difficult to execute a quantum circuit on these processors successfully.

Quantum algorithms, usually represented as quantum circuits, are hardware-agnostic; that is, when described, they do not consider hardware restrictions. To execute such algorithms (quantum circuits) on a quantum processor, they must be modified to fulfill the processor's limitations through a process called quantum circuit mapping. The quantum circuit mapper, which is part of the compiler, is then at the core of the full-stack quantum computing system, connecting algorithms with quantum devices \cite{bandic2022full}. 

Various techniques have been proposed to deal with the mapping of quantum circuits \cite{li2019tackling,  murali2019noise, tannu2019not, li2020towards, zulehner2018efficient, venturelli2019quantum, lao2018mapping, lao2019mapping, herbert2018using}, which differ in approach (exact or heuristic, local or global solution), methodology (e.g. SMT solver \cite{lye2015determining}), cost functions (optimizing number of gates or circuit depth) and performance metrics (e.g. circuit fidelity). These solutions, however, adopt a bottom-up approach, developing mappers specifically for certain quantum processors and technologies. The majority of quantum circuit mapping techniques have mostly focused on hardware properties \cite{lao2021timing,tannu2019not} and only considered a rather limited set of algorithm characteristics such as number of qubits, number of quantum gates, two-qubit gate percentage, and qubit interactions (i.e., what pair of qubits perform a two-qubit gate). In addition to this, when mapping outcomes are analyzed, the focus is on the values of the obtained metrics without further evaluating why some circuits show higher or lower overheads. Some works have already pointed out the importance of including more algorithm features in the mapping process \cite{https://doi.org/10.48550/arxiv.2108.02099}. A more complete and  in-depth profiling of quantum circuits will help to: i) have a deeper understanding on why specific algorithms have higher fidelity than others when being run on a particular processor using a specific mapping technique; ii) to categorize (cluster) quantum circuits based on those parameters and predict the performance of additional circuits with similar properties in terms of mapping-related metrics, without actually running them on a given device; and iii) to develop application-driven and hardware-aware mapping techniques (i.e., mapping techniques tailored for a specific set of algorithms in addition to overcoming hardware constraints) \cite{li2021software,lao20212qan,bandic2022full}. Note that more broadly, this characterization of quantum circuits will be also crucial for defining a meaningful and complete set of quantum benchmarks to evaluate not only quantum circuit mapping techniques but also full-stack quantum computing systems as well as for having a set of algorithm-level metrics to measure system performance \cite{tomesh2022supermarq}.

One of the most stringent quantum hardware constraints that quantum circuit mapping techniques have to deal with, is the limited connectivity of physical qubits, which restricts possible interactions between them. Therefore, in this paper, we propose to extend the profiling of quantum circuits/algorithms by not only extracting `standard' parameters like the number of qubits and gates and percentage of two-qubit gates, but also by performing a deeper analysis of their qubit interaction graphs (i.e., representation of the two-qubit gates or qubit interactions of the circuit). 
By taking input from graph theory and machine learning, we characterize quantum circuits based on their interaction graph metrics (e.g., average shortest path, connectivity, clustering coefficient). We then map those quantum circuits into several quantum processors using a specific quantum circuit mapping technique. In future work, we will also use different quantum circuit mapping configurations, allowing us to evaluate what quantum circuit features impact the circuit mapping performance the most and identify what combination of mapping technique-quantum hardware works better for a given (set of) algorithm(s). Note that this analysis can in the future help in the codesign of algorithm-driven compilation methods and quantum hardware.  

In addition, we present a categorized and as of now the most comprehensive set of quantum algorithms (benchmarks) from various sources and platforms and in different quantum programming languages. Most of the currently existing and used quantum algorithms, synthetically generated and application-based circuits are included in this collection and classified based on different criteria. We are hoping that this algorithms/circuits set will be used for benchmarking quantum computing systems as well as parts of it, such as compilation techniques.

The main contributions of this work are:

\begin{enumerate}
    \item We have performed the first characterization and clustering of quantum circuits that also considers qubit interaction graph parameters in addition to the characteristics related to circuit size (number of gates, number of qubits, amount of two-qubit gates). In-depth profiling and clustering of quantum circuits based on their more structural parameters help to analyze why and when some (families of) quantum algorithms show better performance compared to the rest when being executed on a given quantum processor, as well as which circuit parameters have a higher impact on performance for some hardware-compiler setups. Subsequently, that can also help to predict the mapping performance for additional circuits with similar properties, without actually running them on a given device, and therefore assist in recommending an adequate mapper and hardware configuration to use. Finally, this circuit structural parameters analysis step is crucial for the development of future application-based quantum devices and mappers.
    \item We have found that quantum circuits similarly structured in terms of their interaction graph parameters will have comparable results in terms of circuit fidelity and gate overhead when mapped on the same quantum device and by using the same mapping technique. By running these groups of circuits with different hardware configurations we could make clear suggestions on which group of circuits fits which hardware better.
    \item We provide the so-far most comprehensive collection of quantum benchmarks, open-source and available in most currently used high- or low-level quantum languages. The goal is to help  the quantum community speed up the research process and in the development of a full-stack quantum system by having an easily accessible, all-in-one-place set of benchmarks that can be used for analyzing the performance of existing and future quantum processors and compilation methods.
\end{enumerate}

The paper is organized as follows: Sect. \ref{Sec2} presents a short introduction to full-stack quantum computing systems and an overview of the current state-of-the-art quantum circuit mapping techniques as well as benchmark characterization. Sec. \ref{Sec3} introduces our profiling of quantum algorithms and their clustering based on size and structure. The experimental setup with the details of our benchmark collection is included in Sec. \ref{Sec4}. Sec. \ref{Sec5} showcases the obtained results on how the mapping performance of quantum circuits when run on a specific chip relates to their structural parameters acquired from the analysis of their interaction graphs and their clusters from Sec. \ref{Sec3}. Finally, in Sec. \ref{Sec6} and Sec. \ref{Sec7}, conclusions and future work are presented.

\section{Background and related work}

\label{Sec2}

\subsection{Quantum computers nowadays}
\label{Sec2.1}

\emph{Quantum hardware} has significantly progressed since its inception, and a wide variety of technologies has been developed for implementing qubits like solid-state spins, trapped-ion qubits or superconducting qubits \cite{resch2019quantum}. Hardware characteristics like the number of qubits and gate fidelity are continuously improving. However, current NISQ devices are still immensely resource-constrained and error-prone. They are not able to keep up with the development of promising \emph{quantum algorithms}, that might achieve exponential speed-up, as they lack in size (number of qubits), which is required for the implementation of fault-tolerant and error-corrected techniques. Therefore, it was inevitable to develop a set of algorithms that could be successfully executed on current processors, coming from different fields like quantum physics, chemistry, or machine learning \cite{Bharti_2022}.

\emph{Quantum compilers} act like intermediaries between algorithms (expressed as quantum circuits) and quantum processors. They not only translate high-level programming language  instructions (e.g., library Qiskit given in Python \cite{Qiskit} ) into low-level ones (quantum assembly-like language, e.g., OpenQASM\cite{OpenQASMall}), but are also responsible for making transformations and optimizations of the quantum circuit to best fulfill the quantum hardware requirements. The compiler design and complexity highly depend on the constraints imposed by the hardware and chosen technology. In nearest-neighbor architectures (e.g., 2D array of qubits), the primary constraint is the limited connectivity among qubits. As running two-qubit gates requires that the paired qubits are adjacent on the chip, restricted connectivity can become a huge obstacle. The compiler tries to overcome that and other limitations and helps to successfully execute a quantum circuit on a given quantum device through a process called mapping. Note that the mapping of quantum circuits usually results in a gate and latency overhead that in turn decreases the circuit fidelity. Therefore,  having efficient mapping techniques is crucial in the NISQ era not only to successfully execute quantum algorithms but also for extracting the most out of constrained NISQ devices.

\subsection{Computing with NISQ devices}
\label{Sec2.2}

One of the motivations for building quantum computers in the first place is to run algorithms that solve problems that are intractable for existing classical computers due to limitations in speed and memory. Current NISQ devices can only handle simple algorithms, in terms of the number of qubits and gates and circuit depth, as the presence of noise and limited resources (physical qubits) still constrain them: quantum operations have high error rates and qubits decohere over time resulting in information loss. On top of that, running an algorithm on a NISQ device is not a straightforward process. That is due to hardware constraints that affect the algorithm execution, which can vary between quantum technologies.

\begin{figure}[t!]
	\centering
	\includegraphics[width=\linewidth]{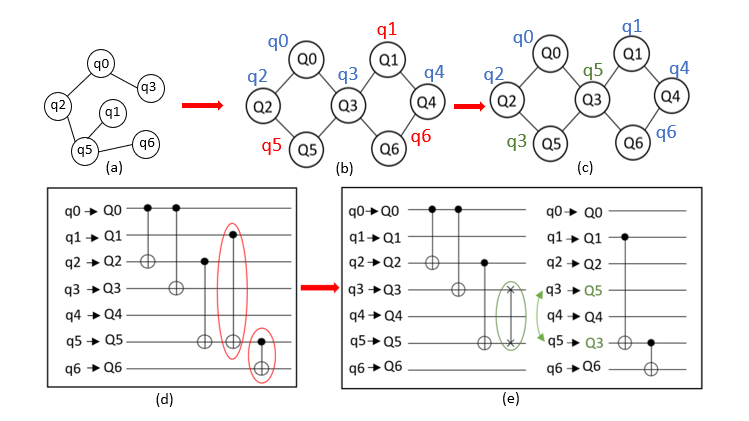}
 	\caption{Running a quantum circuit on a 7-qubit quantum processor. (a) Interaction graph $G_i(V_i, E_i)$ of the circuit shown below; Nodes $V_i$ represent virtual qubits, and edges $E_i$ show interactions between qubits (i.e., 2-qubit gates). 
 	  (b,c) The chip's coupling graph $G_c(V_c, E_c)$; Nodes $V_c$ represent physical qubits, edges $E_c$ show connections on the chip (i.e., possible two-qubit interactions). (d) Circuit qubits ($qi \in V_i$) are mapped onto physical qubits ($Qi \in V_c$). (e) An extra SWAP gate is required to be able to perform all CNOT gates.}
 	\label{fig:mapping1}
\end{figure}

One of the restrictions that affects the execution of a quantum algorithm the most is \emph{(limited) qubit connectivity}. That applies to most technologies, including superconducting qubits and quantum dots, where qubits are arranged in a 2D grid or some other not-fully connected topology, as shown in the top-right part of Fig. \ref{fig:mapping1}, allowing only nearest-neighbor interactions. In order to perform a two-qubit gate in such architecture, the two interacting qubits in the circuit have to be placed in neighboring physical qubits on the chip, which is not always possible (see Fig. \ref{fig:mapping1}: two two-qubit gates between virtual qubits 1 and 5, and 5 and 6 cannot be directly performed because they do not share a physical connection in the coupling graph). Other constraints that have to be considered are: i) \emph{primitive gate set} -- the gates of the circuit to be executed do not always  match the native gate set (supported gates) of the quantum chip. For instance, to run the quantum circuit shown in Fig. \ref{fig:mapping1} on the Surface-17 chip \cite{lao2021timing}, its CNOT gates would have to be decomposed into X and Y rotations and CZ-gate supported by the device; ii) \emph{classical control constraints} -- shared electronics help to scale up quantum systems but may limit parallelization of quantum operations during circuit execution.
The process of accommodating these requirements imposed by the quantum hardware to efficiently execute a quantum algorithm is called \emph{quantum circuit mapping}.

The quantum circuit mapping process consists of the following steps (not mandatory in this order): 1) \emph{Adapting the gate set} of the circuit to the gates supported by the device; 2) \emph{Scheduling} quantum operations (qubit initialization, gates and measurements) of the circuit to leverage its parallelism and therefore shorten the execution time; 3) \emph{Placing virtual qubits} (of the circuit) \emph{onto physical qubits} (on the actual chip) so that the previously mentioned nearest-neighbor two-qubit-gate constraint is satisfied as much as possible during algorithm execution; and 4) \emph{Routing} or exchanging positions of virtual qubits on the chip such that all qubits that could not initially interact become adjacent and perform their corresponding two-qubit gates (Fig. \ref{fig:mapping1}). This is done by inserting additional quantum gates. How routing is performed and which gates are inserted is technology-dependent with various existing methods (SWAPs, Shuttling). Therefore, the resulting after-mapping circuit will in most cases have more gates and a longer execution time than originally. Due to the previously mentioned highly-erroneous quantum operations and qubit decoherence, the overhead in terms of number of gates and circuit depth caused by the mapping should be minimal as it ultimately impacts the algorithm fidelity.

Various approaches have been proposed to solve the circuit mapping problem, each using different methods and strategies. Some solutions are optimal (exact), but work in a brute-force style and are thus only suitable for small circuits \cite{lye2015determining,siraichi2018qubit, zulehner2018efficient}. For larger circuits and to allow for scalability, heuristic solutions are a better fit \cite{wille2016look,guerreschi2018two, li2019tackling, lao2021timing}. Some methods proposed by related works include the use of SMT solvers \cite{lye2015determining,murali2019noise}, greedy heuristic \cite{dousti12min,bahreini2015minlp, li2019tackling,zulehner2018efficient} and machine learning-based algorithms \cite{venturelli2018compiling, herbert2018using, pozzi2020using}. 
These solutions all focus on the `routing' part of the mapper. In addition to this, it is possible to deal with the mapping problem by optimizing its other stages like scheduling \cite{lao2021timing,guerreschi2018two}, gate transformation \cite{Guerreschi2019, itoko2020optimization,tan2021optimal,pozzi2020using} or initial placement \cite{jiang2021quantum,li2020qubit,tannu2019not}. 

\begin{figure}[h!]
    \centering
\subfigure[]{\includegraphics[width=0.5\linewidth]{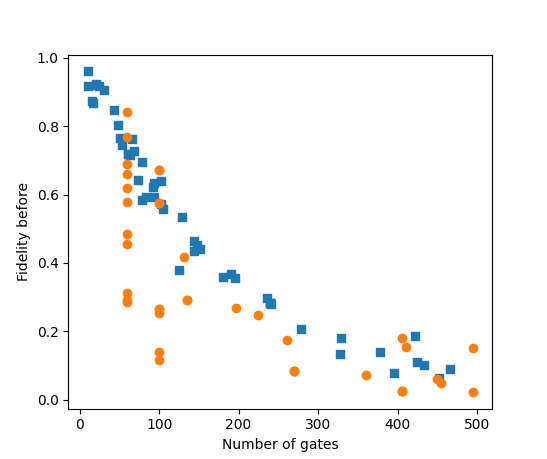}\hspace{-6mm}
\label{fig:fidDecGates}}
\subfigure[]{\includegraphics[width=0.5\linewidth]{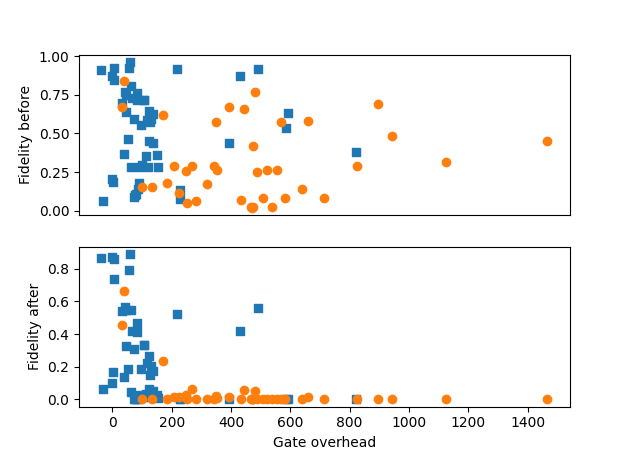}
\label{fig:fidDecdgateo}}
\caption{(a)  Circuit fidelity vs. the number of gates. (b) Gate overhead (\%) and decrease in fidelity. Synthetically generated circuits marked with orange circles and real ones (i.e., quantum algorithms and routines) with blue squares. Here, only circuits with up to 500 gates were used.}
\label{fig:graphs1}
\end{figure}

Different metrics are being used to assess the performance of the quantum circuit mapping technique depending on the cost function: some works have the goal of minimizing the number of gates or gate overhead (e.g., number of additional SWAP gates) \cite{zulehner2018efficient, bandic2020structured, lao2021timing, itoko2020optimization,tan2021optimal,hillmich2021exploiting,lao2018mapping, li2020qubit}, some prioritize low circuit depth or latency (circuit execution time) \cite{zulehner2018efficient,tan2021optimal,lao2018mapping, pozzi2020using, bandic2020structured,lao2021timing} and finally some focus on the success rate of the circuit \cite{jiang2021quantum,blume2020volumetric} and maximizing fidelity \cite{murali2019noise,tan2021optimal,tannu2019not} by also considering the different error rates of the quantum device. Note that the overall goal in the current NISQ era is to maximize the fidelity and success rate of quantum circuits, which currently mostly depends on the gate and circuit depth overhead. Fig. \ref{fig:graphs1} shows the impact of the number of gates and the gate overhead on the circuit fidelity. However, as shown in Fig. \ref{fig:fidDecdgateo}, not all the circuits end up with the same decrease in fidelity for the same or similar gate overhead. Note that the circuit fidelity is close to 0\% for any circuit with more than 500 gates (Fig. \ref{fig:graphs1}a). In addition, a gate overhead of over 200\% after mapping leads, in most cases, to a 100\% fidelity decrease (Fig. \ref{fig:graphs1}b).

These approaches all have in common that they are designed to adapt quantum circuits to the device-specific properties and constraints considering only a reduced set of algorithm properties such as gate and qubit count and two-qubit gate percentage (including qubit interactions). A more in-depth quantum circuit characterization, which for instance could include characteristics of the qubit interaction graph like the number of times each pair of qubits interacts and  the distribution of those interactions among the qubits, and of the quantum instruction dependency graph (i.e., graph that represents the dependencies between gates in the circuit and used for scheduling) is still missing. Looking further into interaction graphs  is very beneficial for the quantum circuit mapping process, as like stated before, the most stringent constraint of current quantum hardware is its limited qubit connectivity. Some authors have already pointed out the importance of including application properties \cite{lubinski2021application,mills2020application,li2020towards, bandic2022full} and considering the characteristics of the qubit interaction graphs for improving the mapping of quantum circuits \cite{9736599,bandic2020structured}. Even in classical computing, we notice that different computing resources are necessarily based on what we use the computers for and which applications are executed. For instance, a dedicated GPU can be used for highly parallelizable processes. Likewise, thorough profiling can help to identify which algorithm characteristics are required to execute it successfully on a given device and vice-versa. The structural properties of quantum circuits can also help understand why specific algorithms show better success rates than others when being run on a particular processor using a specific mapping technique.

\section{Profiling of quantum circuits based on qubit interaction graphs} 
\label{Sec3}

This section provides an overview of the qubit interaction graph-based benchmark profiling and clustering process, emphasizing why this could be meaningful for improving future quantum circuit mapping techniques.

\subsection{On the importance of qubit interaction graphs for quantum circuit mapping}
\label{Sec3.1}

\begin{figure}[h]
	\centering
	\includegraphics[width=0.7\linewidth]{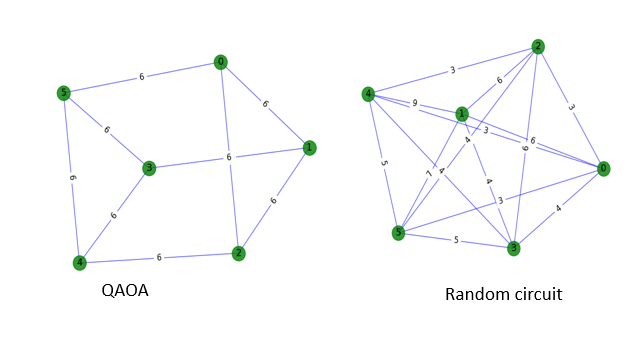}
 	\caption{Interaction graphs of circuits with same size-related parameters: no. of qubits = 6, no. of gates = 456, two-qubit gate percentage = 0.135. } 
 	\label{fig:interaction}
\end{figure}

\emph{Qubit interaction graph} $G(V, E)$ is a graphical representation of the two-qubit gates of a given quantum circuit. It is in general a directed connected graph. Fig. \ref{fig:mapping1} shows an example of a quantum circuit Fig. \ref{fig:mapping1}(d) along with its interaction graph $G_i(V_i, E_i)$ representation Fig. \ref{fig:mapping1}(a). Directed edges $E_i$ represent two-qubit gates and nodes $V_i$ are the qubits that participate in them. Since the direction of edges in most cases doesn't influence the execution of the gates, it is sufficient to perceive the interaction graph as undirected for the mapping problem \cite{prielinger2023aquadratic}. If a circuit comprises multiple two-qubit gates between pairs of qubits, it results in a weighted graph (like in Fig. \ref{fig:interaction}), which shows how often each pair of qubits interacts and how those interactions are distributed among qubits.

This additional information can be leveraged to provide more insights into a circuit structure that is otherwise hidden when only considering standard algorithm parameters such as the number of qubits and gates and two-qubit gate percentage. To illustrate this, Fig. \ref{fig:interaction} shows the interaction graphs of two quantum algorithms, an instance of QAOA and a randomly generated circuit (on the right), which a priori are similar when only characterized in terms of the three common algorithm parameters. What can be noticed is that their qubit interaction graph structure is quite different: the graph of the random circuit is more complex with full connectivity and presents a different distribution of the interactions between qubits, that is, of the weights. This will result in more routing and, therefore, higher overhead, unless we indeed have a fully connected coupling graph of the processor (Sec. 5). 

This shows the importance of quantum circuit structure when developing mapping techniques and the necessity of characterizing the circuits in terms of their qubit interaction graphs.
A few works have already pointed out how interaction graph along with quantum instruction dependency graph can be used as a baseline for designing better mapping techniques \cite{li2019tackling,lao2021timing,baker2020time,bandic2023mapping}. In those works, gate dependency graphs are used as core information for scheduling optimization and look-ahead techniques, whereas interaction graphs are usually only used for the initial placement of qubits of the routing procedure.  Considering that the primary constraint affecting the fidelity of the circuit execution is nearest-neighbor connectivity required for performing two-qubit gates, it would be valuable to know in advance how they are distributed among qubits and not only their quantity.

In this paper, we perform profiling of quantum circuits by focusing on interaction-graph properties and their relation to quantum circuit mapping. To that purpose, we took input from graph theory and analyzed qubit interaction graphs based on metrics described in \cite{hernandez2011classification} with a focus on those that are relevant to the mapping problem.

Quantum circuit profiling in our work consists of the following steps:

\begin{enumerate}
    \item \textbf{Benchmark collection} -- collecting benchmarks (quantum circuits) from various sources, translating them to the same quantum language and extracting their interaction graphs (Sec. \ref{Sec4}).
    \item \textbf{Parameter selection and extraction} -- choosing and extracting graph-theory-based parameters from the qubit interaction graph that are relevant to the mapping of quantum circuits.
    \item \textbf{Benchmark clustering} -- clustering benchmarks based on their size- and interaction graph-related parameters.
\end{enumerate}

After performing these steps, we compiled the quantum circuits using OpenQL \cite{khammassi2021openql} and analyzed the relation between their performance and extracted parameters, as well as clusters (Sec. \ref{Sec4} and Sec. \ref{Sec5}).

\subsection{Parameter selection for quantum algorithm profiling}

There exists a vast amount of metrics used for describing graphs, which can be classified into different groups and classes. However, not all of these metrics are relevant to our goal in terms of qubit interaction graph analysis. After thoroughly investigating all metrics described in \cite{hernandez2011classification}, we chose those that are key for the circuit mapping problem. These metrics, when calculated from the qubit interaction graphs, should represent features of quantum circuits that have a correlation with the mapping performance metrics (e.g. number of SWAPS). For instance, the node degree distribution is a relevant metric as it defines the connectivity of the graph (i.e. density of qubit interactions). The more connected the graph, the higher the node degrees. In case there is an all-to-all connected interaction graph all degrees would be $n-1, (n$ being the number of qubits) and that graph would be more challenging to map onto limited connectivity device topologies, which would result in the insertion of a higher number of additional SWAP gates. Table \ref{tbl:metr1} shows the selected metrics subset and how they relate to the quantum circuit mapping process.

\begin{table}[h!]
	\centering
	\includegraphics[width=\linewidth]{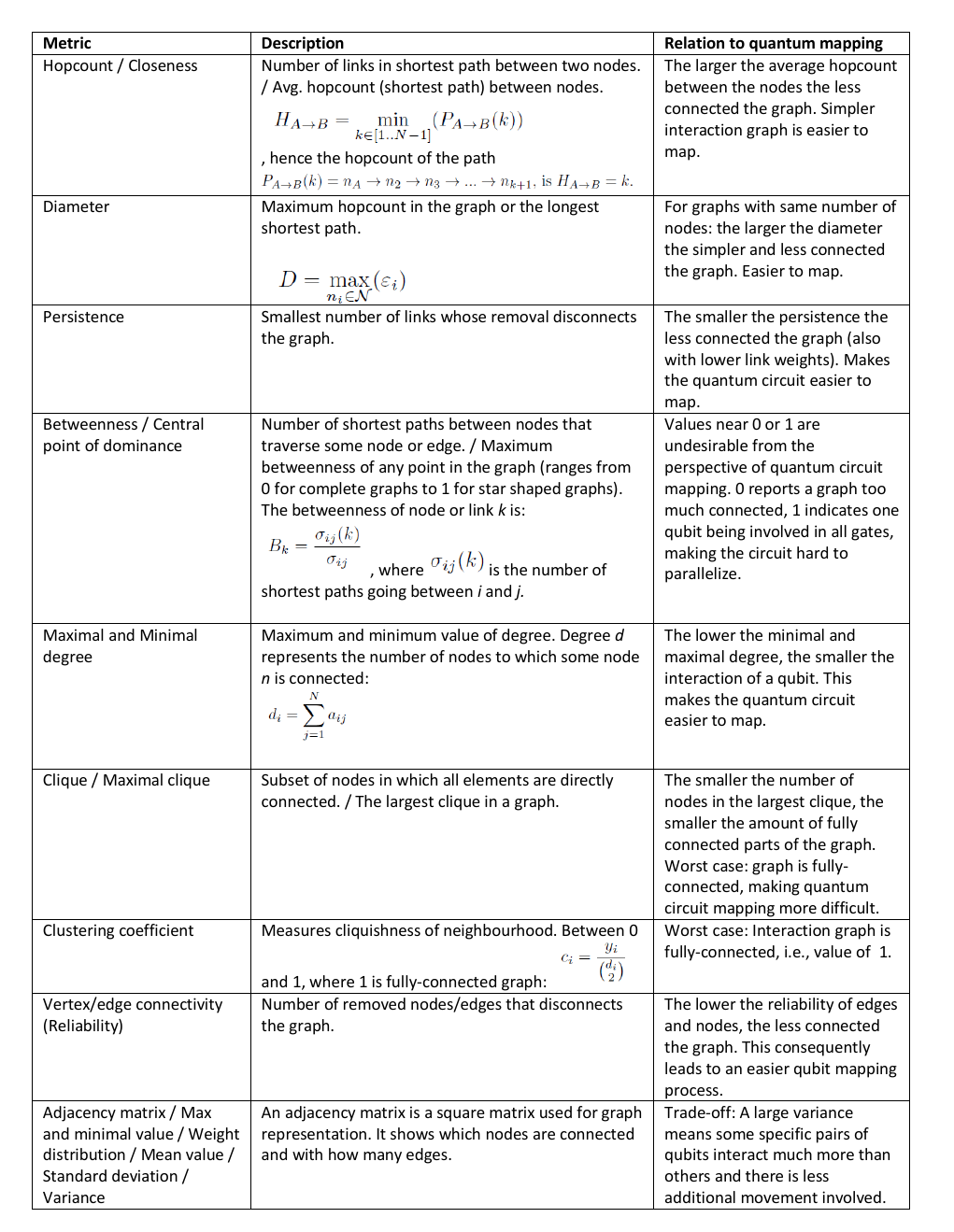}
 	\caption{Selected metrics for characterizing interaction graphs and their relation to the quantum circuit mapping.}
 	\label{tbl:metr1}
\end{table}

\begin{figure}[h]
	\centering
	\includegraphics[width=\linewidth]{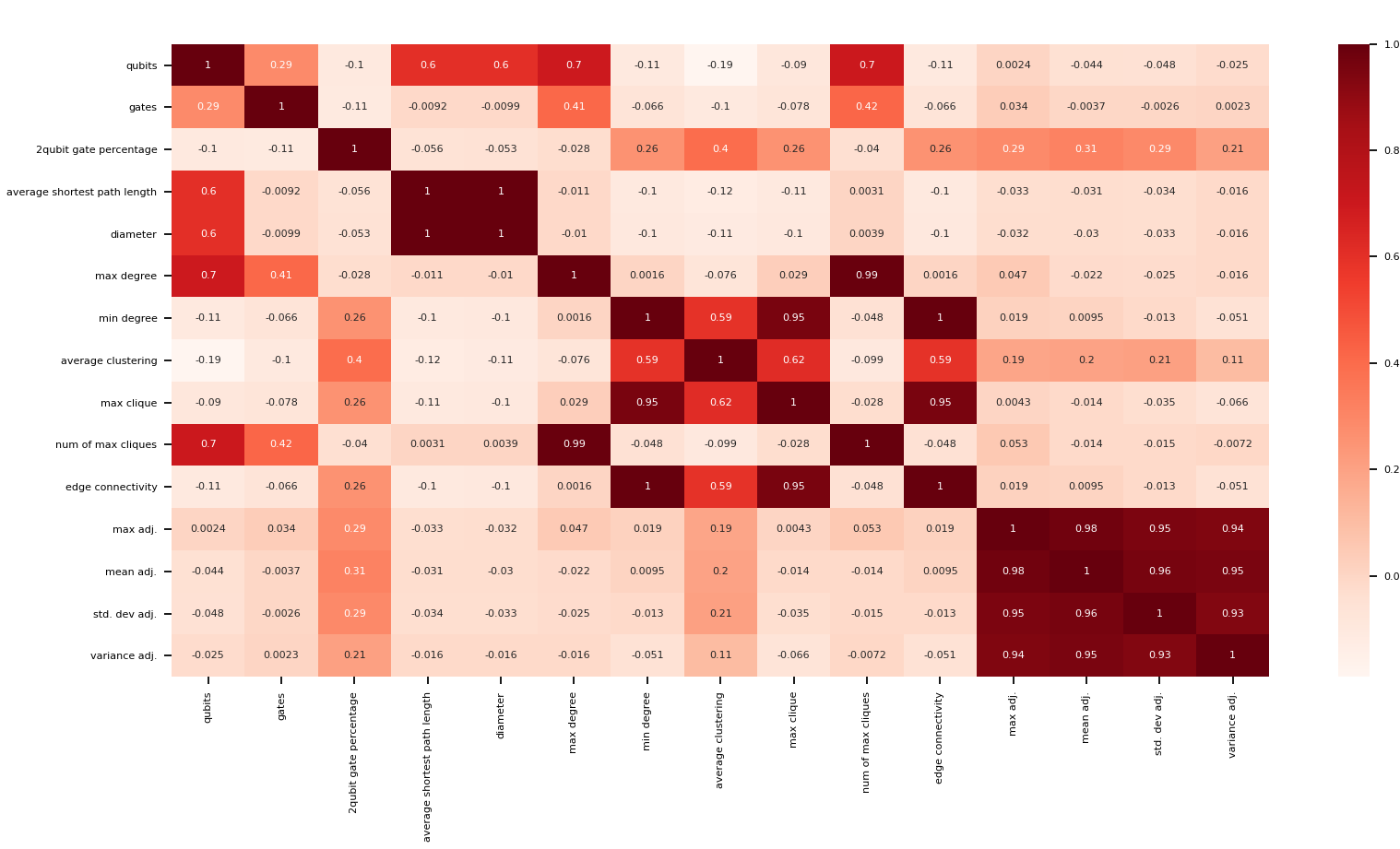}
	\caption{Heatmap of a Pearson correlation matrix for quantum circuit and interaction graph metrics selected for mapping.}
 	\label{fig:heatmap}
\end{figure}

We noticed, however, that a large amount of these metrics are correlated, i.e., they scale in the same manner. Therefore, the parameter space was reduced by using a Pearson correlation matrix as shown in Fig. \ref{fig:heatmap} (-1/1 meaning maximally-correlated, 0 meaning not correlated) \cite{freedman2007statistics}. For instance, note that a minimal node degree of a graph strongly relates to maximal clique and edge connectivity, so in that case just using one of the parameters, instead of all three, is sufficient. This method allowed us to reduce our previous metric set to: average shortest path (average hopcount), maximal and minimal node degree, and adjacency matrix (interaction graph edge-weight distribution) standard deviation. These metrics and the common circuit parameters can be used to cluster quantum circuits. It is expected that quantum algorithms with similar properties should show similar  performance when run on specific chips using a given mapping strategy. 

\subsection{Clustering benchmarks outcomes and evaluation}

\begin{figure}[h]
	\includegraphics[width=\linewidth]{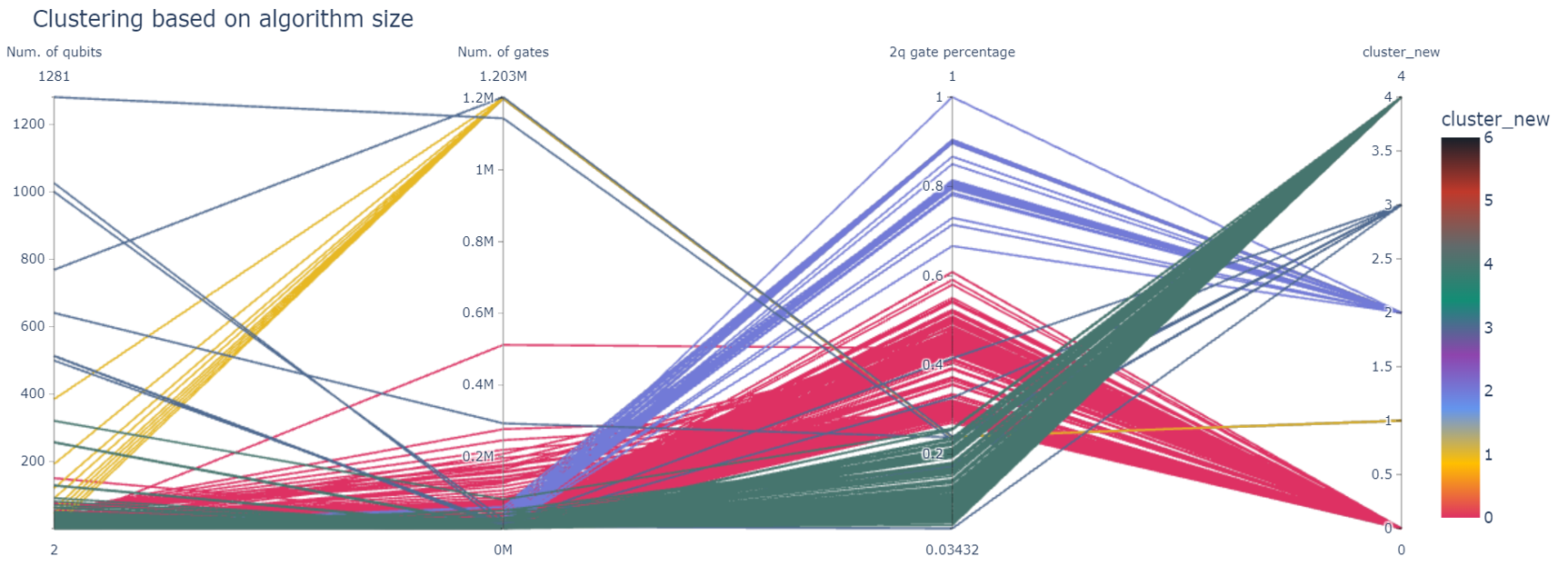}
	\caption{Clustering of quantum algorithms based on size-related parameters.}
 	\label{fig:kmeans}
\end{figure}

As mentioned earlier, one of our goals is to find structural similarities among quantum circuits and create some sort of `circuit families', whose elements (quantum circuits) will show similar compilation behavior and require similar hardware resources. The two criteria we have used for clustering benchmarks are properties based on circuit size and qubit interaction graph. Note that we performed a two-step clustering: circuits were first clustered based on size parameters (number of qubits and gates and percentage of two-qubit gates) and then on qubit interaction graph metrics.  The reason behind this was to avoid the former to become the most significant criteria of our clustering algorithm. Fig. \ref{fig:kmeans} shows the five clusters (different colors) in which a set of 300 selected benchmarks (Sec. \ref{Sec4}) have been divided by using the kmeans \cite{1056489} clustering algorithm. Benchmarks are represented as lines in this parallel-coordinates plot. The x-axes contains a list of three different parameters with their values shown in y-axes. 

\begin{figure}[h!]
	\centering
	\includegraphics[width= \linewidth]{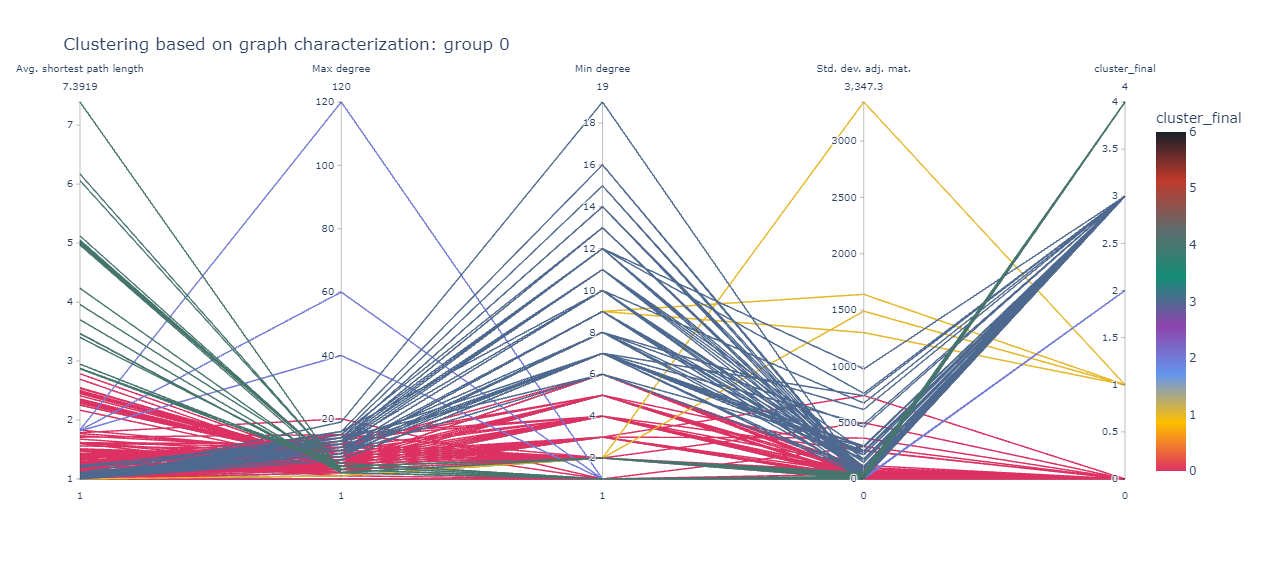}
	\caption{Sub-clustering of quantum algorithms of cluster 0 (Fig. \ref{fig:kmeans}) based on interaction graph parameters.}
 	\label{fig:kmeans2}
\end{figure}

Each of the five size-related clusters can then be further divided into sub-clusters based on previously explained graph parameters: average shortest path length, maximal and minimal degree and adjacency matrix standard deviation. In this case, we have again selected the kmeans algorithm among several others by evaluating different methods and parameter setups with the silhouette coefficient method \cite{rousseeuw1987silhouettes}. In Figure \ref{fig:kmeans2}, an example when one of the size-parameters-based clusters (cluster 0 from Fig. \ref{fig:kmeans}) is divided into sub-clusters based on the interaction graph parameters. It is also pretty straightforward for additional future circuits to be assigned to a specific cluster (size- and interaction graph-based) as each of the clusters and sub-clusters covers the specific range of combinations of parameters (e.g. cluster 4 in Fig. \ref{fig:kmeans} covers benchmarks with less than 25\% percentage of two-qubit gates, cluster 3 in Fig. \ref{fig:kmeans2} covers the highest minimal degree values (over 6)). Those circuits should then have similar expected fidelity and gate overhead outcomes as the other circuits in that cluster. How exactly do the mapping performance metrics correlate with our clusters from Fig. \ref{fig:kmeans2}, and the possible reason for that will be described in the next sections.

\section{Experimental setup}
\label{Sec4}

This section describes all the necessary elements for performing our experiments: i) our newly created benchmarks collection \cite{qbench} and a subset used in this paper; ii) the OpenQL compiler with its Qmap mapper \cite{lao2021timing} and Surface-97 platform, IBM Rochester and Aspen-16 configuration files and iii) chosen set of metrics for evaluating the performance of the quantum circuit mapping technique.

\subsection{Quantum benchmarks collection and classification}

\label{Sec4.1}

\begin{figure}[h]
	\centering
	\includegraphics[width=0.8\linewidth]{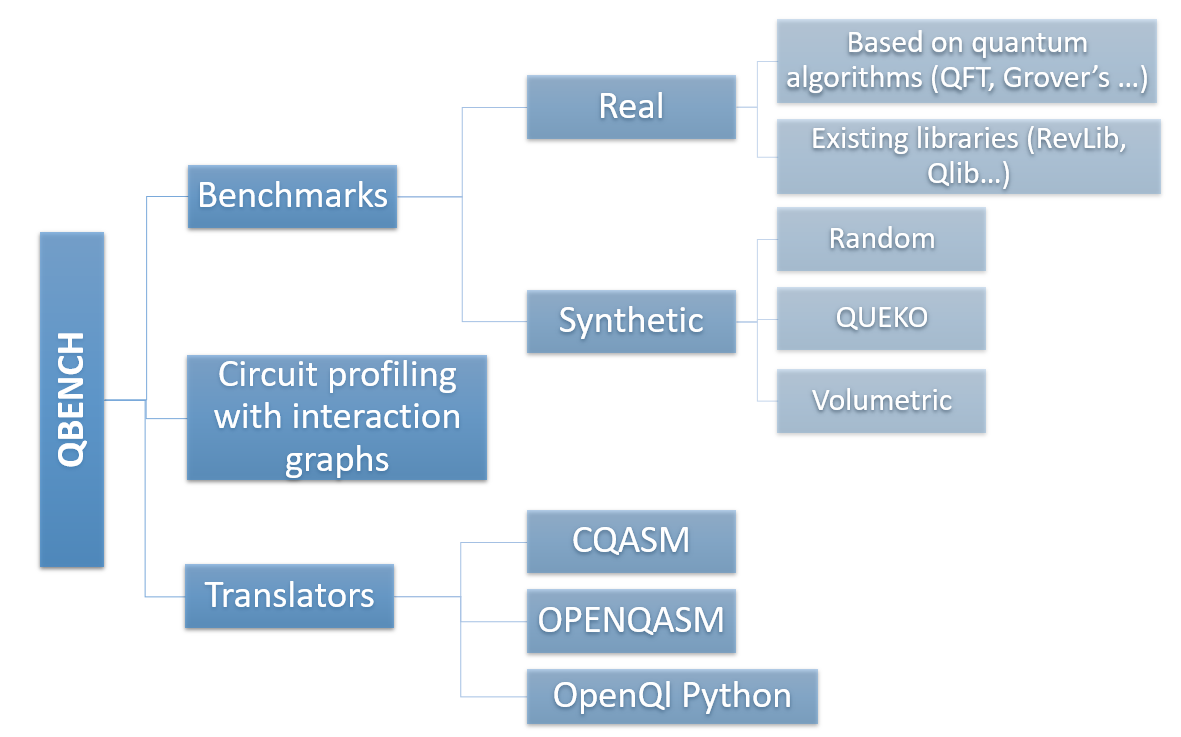}
	\caption{Overview of our QBench repository }
 	\label{fig:benchcol}
\end{figure}

The fast development of quantum computing systems dictates the necessity for an all-including and standardized benchmark suite that can serve to test  quantum devices as well as compilation techniques, and in general, any part(s) of the full stack. To address this issue, we collected various types of quantum circuits used as benchmarks from a large number of sources \cite{li2020qasmbench,OpenQASMall,Queko,ZulehnerCircuits, moller2020cross,DiogoRandomCirc,MicrosoftQuantum,Qiskit,QuTechPython,cirq_developers_2022_6599601, 1608.03355,sivarajah2020t,cross2018ibm, 196882c3786049668004cfb4f9748f56, wille2008revlib} written in and translated to different available high- and low-level languages. An overview of our open-source benchmark suite called QBench \cite{qbench} is shown in Fig. \ref{fig:benchcol}.

Benchmarks are first divided into two high-level groups: real  vs. synthetic quantum circuits. The first ones are then further split into two categories depending on whether they are based on quantum algorithms or are simple reversible arithmetic circuits. In the second group, we can find three different subgroups based on how they are generated. According to \cite{nielsen2002quantum}, currently used benchmarks based on \textbf{real algorithms} (QFT, search algorithms, application-based algorithms) are the ones that are of the highest importance when measuring the performance of all future quantum systems as they are scalable, meaningful and can show the advantage in quantum systems comparing to classical counterparts \cite{tomesh2022supermarq}. For the current NISQ era, however, there is a need for benchmark libraries like RevLib \cite{wille2008revlib} that are within the domain of reversible and quantum circuit design. \textbf{Synthetic benchmarks} represent the group of randomly generated quantum circuits, which provide a larger variety in terms of their parameters (e.g., number of qubits, gates, two-qubit gate ratio, circuit depth), and are mainly used to test the performance of quantum devices and explore their computational power to the fullest. For this paper, we mainly focused on: i) randomly generated quantum circuits that are created by uniformly randomly choosing single- and two-qubit gates from a predefined set and then applying them on arbitrarily chosen qubits or qubit pairs in the circuit \cite{DiogoRandomCirc}; ii) QUEKO circuits \cite{Queko}, which are designed to be optimal for specific devices (e.g., with optimal depth) and iii) Quantum volume square circuit \cite{cross2019validating} that is used in general for benchmarking quantum system architectures. A summary of all the real-algorithm-based or synthetic circuits that are part of our benchmark set can be found in \cite{qbench}.

Benchmarks in our set are also classified  based on their size (large-, middle- and small-scale and parameterized ones) and on the higher- or lower-level language they are written in \cite{qbench}. \textbf{Note that a parameterized (scalable) version} of the circuits  allows  the creation of new circuits of a desired size, which will be very meaningful for future quantum systems \cite{tomesh2022supermarq}. Furthermore, \textbf{different translators} from one quantum language to another, \textbf{interaction graphs} and \textbf{interaction graph-based profiling} are also part of this benchmarks suite.

For our experiments, we selected a subset of 300 benchmarks from QBench covering different types (previously described in this section) and qubit number ranges (2-1281 qubits for clustering, 3-54 qubits for mapping experiments).

 Note that this benchmark set is to become open source not only for other researchers to use it for the future development of quantum systems, but also for others to participate in its future extensions. There will always be new benchmarks that can be added or quantum languages to translate the current benchmarks to, as we are in the era where we witness a continuous development of new quantum algorithms, compilers, simulators, and programming languages.

\subsection{Quantum compiler and targeted quantum devices}

\label{Sec4.2}

\begin{figure}[h!]
    \centering
    \centerline{
\subfigure[]{\includegraphics[width=0.3\linewidth]{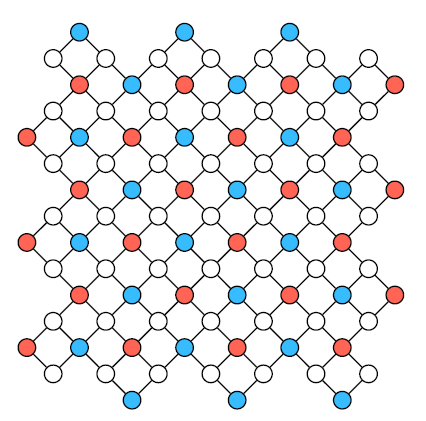}
\label{fig:surf97}}
\subfigure[]{\includegraphics[width=0.3\linewidth]{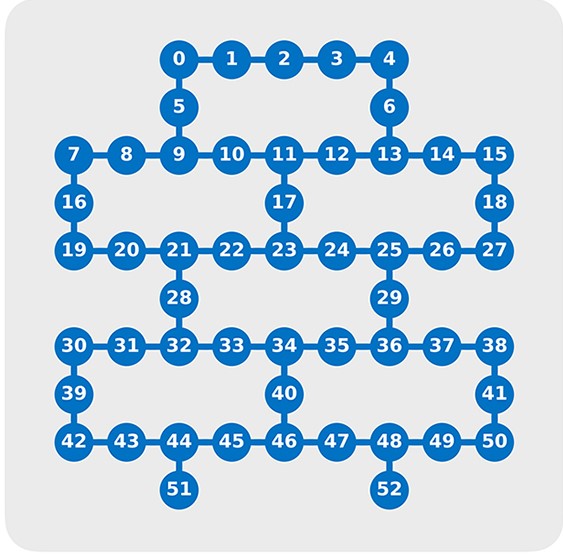}
\label{fig:ibm53}}
\subfigure[]{\includegraphics[width=0.4\linewidth]{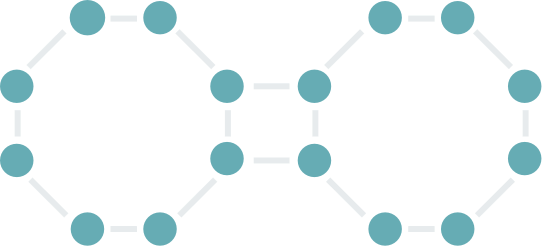}
\label{fig:asp16}}
}
\caption{Tologies of the quantum architectures used for experiments: a) Surface-97; b) IBM Rochester and c) Rigetti 16-q Aspen.  Figures taken from: \cite{overwater2022neural,IBM,Rigetti}.}
\label{fig:topologies}
\end{figure}

To analyze how the previously shown clusters of circuits (Sec. \ref{Sec3}) relates with their after-mapping outcomes, we compiled the 300 selected quantum circuits using as target quantum processor an extended 97-qubit version of the Surface-17 chip (like in Fig. \ref{fig:surf97}). Surface-17 is a quantum processor with a surface code architecture \cite{lao2021timing}, designed to be easily scalable. The device characteristics and all its constraints, are included in a configuration file, which is then used as input for the compiler OpenQL \cite{khammassi2021openql}. The configuration file of our chosen back-end includes information like error rates, primitive gate set, gate-decomposition rules and processor qubit topology/connectivity. In addition to this, and in order to compare the performance of the mapper for different groups of circuits, we performed the same experiments for two more quantum processors: the IBM Rochester and the Rigetti 16q-Aspen chips that are shown in Figs. \ref{fig:ibm53} and \ref{fig:asp16}, respectively. We selected these device configurations because they are currently commonly used in other research on quantum circuit mapping and provide realistic and different connectivity patterns in their coupling graphs. Note that in our experiments we do not execute the quantum circuits on actual devices but instead they are just mapped into the different quantum processors; that is, their hardware constraints are considered in the compiling process.

At the core of the OpenQL compiler is its Qmap mapper, which has many options and strategies allowing to create a sort of custom-made compilation technique. The Qmap quantum circuit mapper considers several types of hardware constraints: limited connectivity, primitive gate set and restrictions derived from classical control electronics. It supports several options for circuit optimization, routing, initial placement as well as scheduling. In addition, it outputs  different circuit mapping performance metrics such as the number of additional gates and circuit latency. The routing strategy we opted for was MinExtend \cite{lao2021timing}, which, among other features, includes looking back to previously mapped gates and strives to minimally extend the latency of the circuit. It also includes different but common gate transformation and optimization strategies such as gate cancellation or commutation.

\subsection{Metrics}

\label{Sec4.3}

The most commonly used metrics for quantum circuit mapper evaluations are the number of added SWAPs, circuit depth and fidelity/reliability. In our case, we have used the additional gates and extended depth information retrieved from the compiler to calculate the following metrics:

\begin{enumerate}
    \item \textbf{Gate overhead} is calculated as:
    
    $G_{overhead} = \frac{(G_{after} - G_{before)}}{G_{before}}$, where $G_{before}$ and $G_{after}$ represent the number of gates before and after compilation. 
    \item \textbf{Latency overhead} is defined as:
    
    $L_{overhead} = \frac{(L_{after} - L_{before})}{L_{before}}$, where $L_{before}$ and $L_{after}$ represent the circuit latency before and after compilation. Latency is calculated as the number of cycles of the circuit, which also considers variations in gate duration, making it different from circuit depth in which all gates are considered to take one time-step. 
    \item \textbf{Circuit fidelity} is defined as the product of error rates of the gates in the circuit. When mapping a circuit, the main goal is to maximize this metric \cite{murali2019full,nishio2020extracting}. We assumed that all one-qubit and two-qubit gates have the same error rates, respectively, for which we used average values of the Starmon-5 chip \cite{qinspire}.
    \item \textbf{Fidelity decrease} is calculated as: 
    $F_{decrease} = \frac{(F_{before} - F_{after})}{F_{before}}$, 
    where $F_{before}$ and $F_{after}$ represent the circuit fidelity before and after compilation.  
\end{enumerate}

In the following section, we will discuss the relation of the structural parameters of circuits with the above-stated obtained metrics after mapping them into the Surface-97, IBM Rochester and Rigetti Aspen-16 devices.

\section{Results}
\label{Sec5}

\subsection{Mapping the circuits to Surface-97 chip architecture}

\label{Sec5.1}

\begin{figure}[h!]
	\centering
	\includegraphics[width=0.9\linewidth]{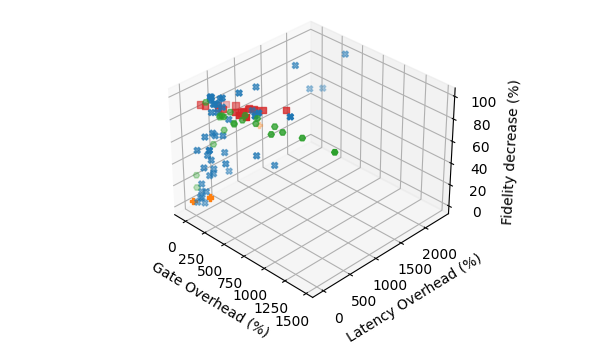}
	\caption{Mapping performance metrics: gate overhead, latency overhead and fidelity decrease (all in \%) for all groups of benchmarks. We differentiate i) synthetic circuits: randomly generated circuits (hexagons) and QUEKO circuits \cite{Queko} (squares) and ii) real, algorithm-based circuits: simpler arithmetic circuits ('x') and circuits based on quantum algorithms ('+') (e.g., QFT or Grover's algorithm, see Sec. \ref{Sec4}). Only circuits with up to 500 gates are shown.}
 	\label{fig:allmetrics}
\end{figure}

In this section, we evaluate and compare the mapping outcomes of our selected circuits and analyze how the circuit parameters impact the results. Additionally, we compare the performance of different clusters of circuits when using the same mapping technique and processor design (Surface-97).

As previously shown in Sec. \ref{Sec2} (Fig. \ref{fig:graphs1}), the gate overhead and circuit fidelity decrease is, on average, higher for our type of synthetic (randomly generated) circuits  than for those based on real algorithms, even when they are in the same range of size.\footnote{The details on how much the fidelity dropped for each benchmark and how much it differs between the two groups are shown in Fig. \ref{fig:barchartfidelity} in Appendices. 
Furthermore, as shown in Fig. \ref{fig:allmetrics}, these two groups of circuits (real and synthetic) are further divided into a total of four differently-structured groups that include: randomly generated circuits, QUEKO benchmarks, quantum algorithm-based circuits, and reversible arithmetic circuits.  Note in Fig. \ref{fig:allmetrics} the difference between these groups in terms of three defined mapping performance metrics. Reversible arithmetic circuits showed on average the lowest gate overhead ($\sim120\%$) and therefore decrease in fidelity. Randomly generated circuits have on average the best latency overhead ($\sim88\%$). To give an example, QUEKO circuits show an average gate overhead of $\sim348\%$, latency overhead of $\sim153\%$ and fidelity decrease of nearly 100\%. This all clearly shows the importance of including the structure of the quantum circuit in the mapping process, and leads us to using that information to our advantage when choosing an appropriate pair of device and mapping technique. 
}

\begin{figure}[h!]
    \centering
    \centerline{
\subfigure[]{\includegraphics[width=0.5\linewidth]{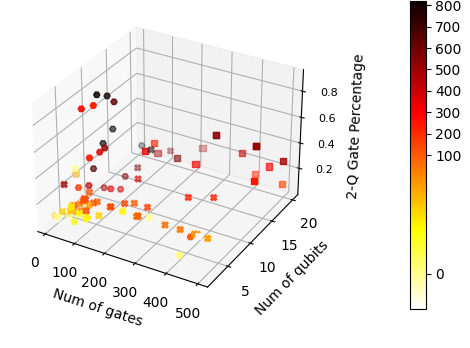}
\label{fig:gateo}}
\subfigure[]{\includegraphics[width=0.5 \linewidth]{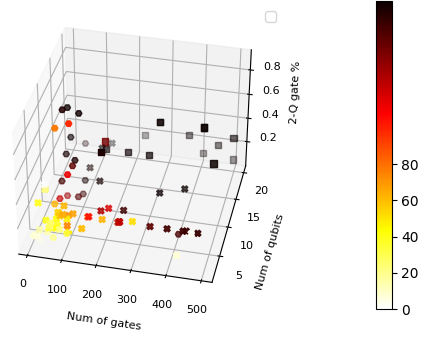}
\label{fig:fidDec}}}
\caption{(a) Gate overhead  and (b)  fidelity decrease in \% (color bar) vs. size-related parameters: number of qubits, number of gates and two-qubit gate percentage. We differentiate i) synthetic circuits: randomly generated circuits (hexagons) and QUEKO circuits \cite{Queko} (squares) and ii) real, algorithm-based circuits: simpler arithmetic circuits ('x') and circuits based on quantum algorithms ('+').}
\label{fig:overhead}
\end{figure}

Subsequent to this, we unveil how size-related parameters: number of qubits, number of gates and two-qubit gate percentage relate to gate overhead and fidelity decrease, respectively as shown in Figure \ref{fig:overhead}. Each point in the graphs represents a benchmark mapped to the Surface-97 processor and just like in Fig. \ref{fig:allmetrics}, different groups of benchmarks are shown using different symbols and in the same way. 
In this case, we only considered circuits with up to 500 gates, as all those above that threshold had negligible fidelity even before mapping. Note that these three mentioned parameters are correlated with the mapping results of the circuits on the chip: the closer the points in graphs are to 0 in all axes simultaneously, the lower the overhead and fidelity decrease. Another point that can be made from these figures is that synthetic circuits (QUEKO and random circuits) perform in this setup, on average, worse than the algorithm-based circuits in terms of after-mapping fidelity and gate overhead (just like in Fig. \ref{fig:allmetrics}).

\begin{figure}[h!]
    \centering
    \centerline{
\subfigure[]{\includegraphics[width=0.5\linewidth]{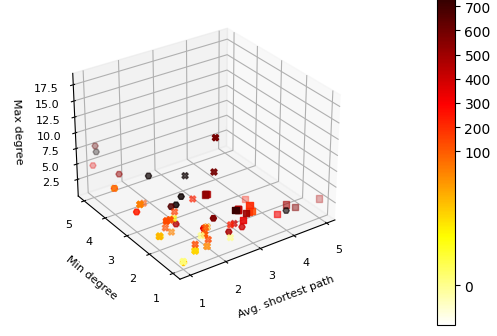}
\label{fig:gateo1}}
\subfigure[]{\includegraphics[width=0.5\linewidth]{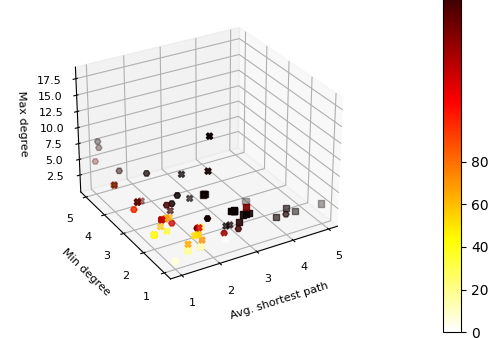}
\label{fig:fidDec1}}}
\caption{(a) Gate overhead  and (b)  fidelity decrease in \% (colorbar) vs. interaction graph-related parameters: minimal node degree, maximal node degree and average shortest path.}
\label{fig:overhead1}
\end{figure}

\begin{figure}[h!]
\centering
\subfigure[]{\includegraphics[width=0.8\linewidth]{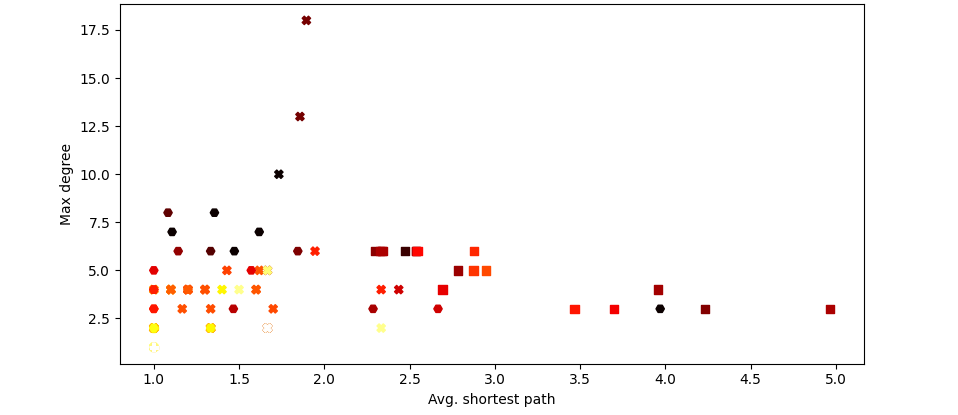}
\label{fig:interparams1}}
\subfigure[]{\includegraphics[width=0.8\linewidth]{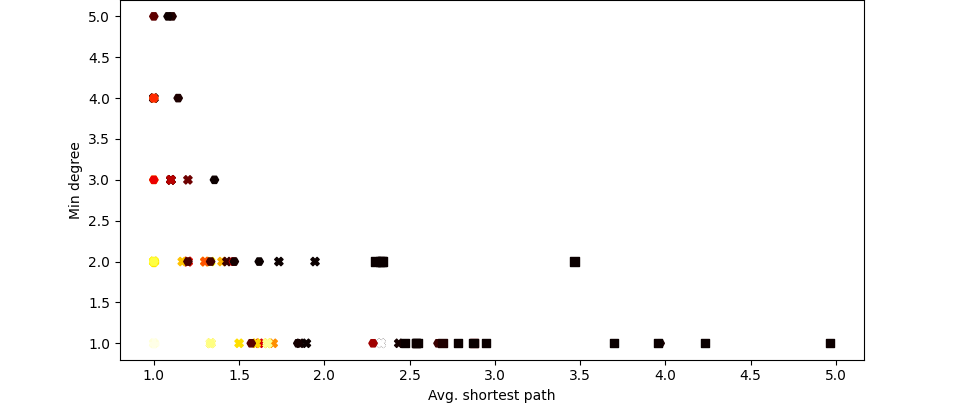}
\label{fig:interparams3}}
    \caption{2D plots of the graphs shown in Fig. \ref{fig:overhead}: (a) interaction graph-based metrics vs. gate overhead (color) and (b) interaction graph-based metrics vs. fidelity decrease (color).}
\label{fig:interparams}
\end{figure}

We have noticed earlier (Sec. \ref{Sec2}) that the size of a circuit, even though an important feature, is not the only reason why some circuits have lower after-mapping overheads than others. Fig. \ref{fig:overhead1} shows how the parameters minimal degree, maximal degree, and average shortest path of the interaction graph influence fidelity and gate overhead of circuits. As observed before,  the closer the points in graphs are to 0 in all axes simultaneously, the lower the overhead and fidelity decrease. The graph shows a strong correlation of both the increase in gate overhead (Fig. \ref{fig:gateo1}) and fidelity decrease (Fig. \ref{fig:fidDec1}) with the increase in maximal and  minimal node degree and average shortest path. 2D cuts of Fig. \ref{fig:overhead1} are shown in Fig. \ref{fig:interparams} for a better visualization. The following observations can be made: 1) the higher all three circuit parameters, average shortest path, minimal and maximal node degree are simultaneous, the higher the gate overhead (Fig. \ref{fig:interparams1}) and fidelity decrease \ref{fig:interparams3}). This means the fidelity is the highest and overhead the lowest if all three circuit parameters are close to 0. 2) Some patterns for circuits belonging to the same group can be observed based on how they are created. For instance, QUEKO circuits (squares) have a high average shortest path ($\sim3$), random circuits (hexagons) have a high average node degree ($\sim8$), whereas RevLib and algorithm-based circuits (x in graph) have on average low values of the same parameters ($\sim1.5$ for average shortest path and $\sim4.5$ for node degree).

\begin{figure}[h]
\centering    
\subfigure[]{\includegraphics[width=0.9\linewidth]{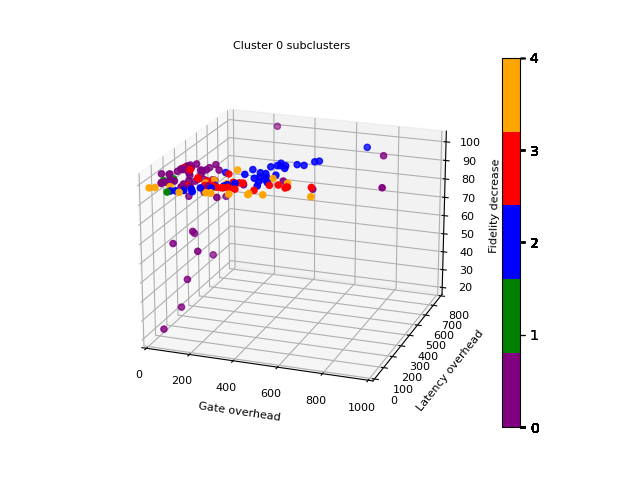}
\label{fig:clusterresultsall}}
\subfigure[]{\includegraphics[width=0.9\linewidth]{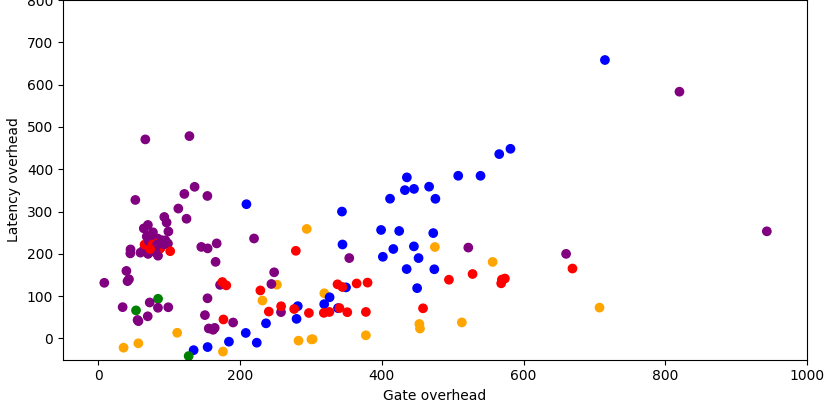}
\label{fig:clusterresults2D}}
\caption{Relation of clusters of circuits (that are shown in Fig. \ref{fig:kmeans2}) with the parameters of their interaction graphs: a) Gate and latency overhead and fidelity decrease and b) Gate and latency overhead}
\label{fig:clusterresults}
\end{figure}
In Sec. \ref{Sec3}, quantum circuits have been clustered based on size and interaction graph parameters. In Fig. \ref{fig:clusterresults}, we can see how the clusters based on interaction graph similarity (example shown in Fig. \ref{fig:kmeans2}) relate to the mapping performance metrics gate overhead, latency overhead, and fidelity decrease. As mentioned in Sec. \ref{Sec4}, the lower these metrics are, the better the mapping performance. One can notice that circuits belonging to Cluster 0 outperform other circuits in terms of gate overhead and fidelity decrease (up to 200\% for gate overhead, and an average of $\sim89\%$ for fidelity decrease), whereas clusters 3 and 4 show the best performance in terms of latency (up to $\sim150\%$). What we can further conclude when comparing Fig. \ref{fig:allmetrics} and Fig. \ref{fig:clusterresultsall}, is that clusters mostly consist of benchmarks of the same type: cluster 0 mostly has real circuits, clusters 3 random ones, and cluster 2 QUEKO circuits. That shows, for instance, that real quantum circuits, especially those from cluster 0, present some pattern in the structure that is easier to map without requiring too many additional gates. Finally, Fig. \ref{fig:clusterresults2D}, which represents a 2D cut of Fig. \ref{fig:clusterresultsall}, clearly shows differences in the range of gate and latency overhead for different clusters. For instance, clusters 3 and 4 have almost constant circuit latency overhead, on average lower than for other clusters, whereas circuits in cluster 0 have low and similar gate overhead. Gate overhead values of cluster 2 scale linearly with latency overhead.

\subsection{Quantum chip topology as one rationale behind results}

\begin{figure}[h!]
    \centering
    \centerline{
\subfigure[]{\includegraphics[width=0.5\linewidth]{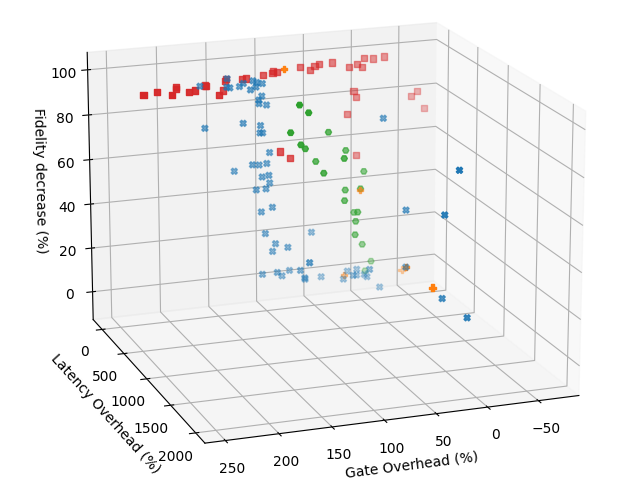}
\label{fig:ibm}}
\subfigure[]{\includegraphics[width=0.5\linewidth]{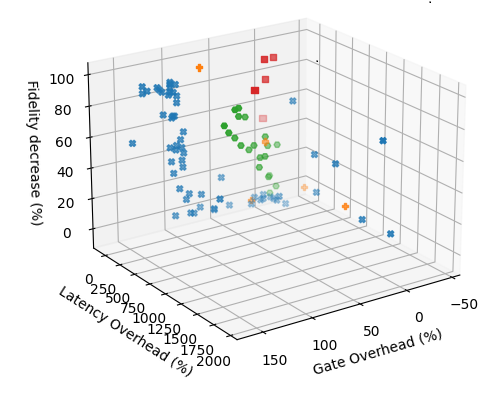}
\label{fig:aspen}}
}
    \caption{Results of the circuit compilation when mapping different quantum  circuits (Random, QUEKO, Reversible arithmetic circuits - RevLib, Quantum-algorithm based circuits) to the IBM Rochester (a) and Aspen-16 (b) device toplogies using the MinExtend mapper}
\label{fig:allmetrics_ibm_aspen}
\end{figure}

\begin{figure}[h!]
	\centering
\includegraphics[width=\linewidth]{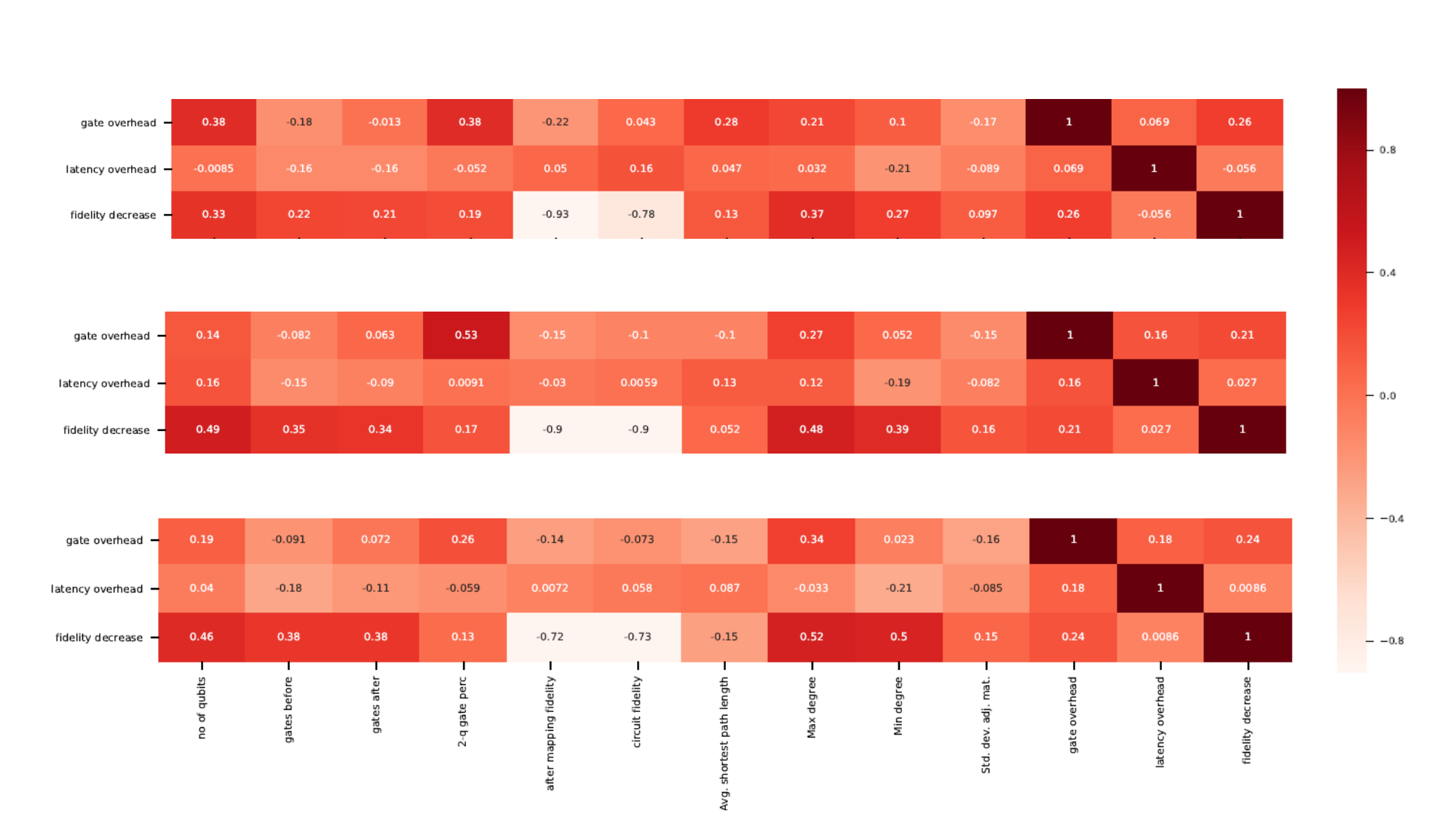}
	\caption{Correlation matrices showing correlations of mapping performance metrics (gate overhead, latency overhead, fidelity decrease) with extracted metrics of the circuit for the three device configurations: Surface-97, IBM Rochester and Rigetti Aspen 16-q (top-down).}
 	\label{fig:corrmetr}
\end{figure}

To further look into the reasoning behind the relation between quantum circuit parameters and mapping performance metrics, we first into the device topology. Thus, we map the same groups of circuits on two additional quantum platforms.: the IBM Rochester and Aspen-16 quantum devices (Fig. \ref{fig:topologies}). The outcomes are shown in Figures \ref{fig:allmetrics_ibm_aspen} and \ref{fig:corrmetr}. Fig. \ref{fig:corrmetr} showcases detailed information on how much each structural parameter influences the three mapping performance metrics: gate overhead, latency overhead, and fidelity decrease for all three device configurations. In Fig. \ref{fig:ibm_aspen} (see Appendix), additional details can be found.

From the figures, we can derive the following: 

i) Different groups of benchmarks based on their origin and structure perform differently when executed on different device topologies. The main value of the figures comes from the fact that we can clearly choose a preferred quantum processor topology for each of the benchmark groups (e.g., Surface-97 is preferred for arithmetic reversible circuits, whereas IBM Rochester might be chosen for random ones, as shown in Figures \ref{fig:allmetrics} and      \ref{fig:allmetrics_ibm_aspen}).

ii) The impact of structural parameters on the results varies depending on the topology. For example, in the case of the two new topologies, the number of qubits was not as strongly correlated with gate overhead, whereas the degree of the graph played a more significant role. The correlation matrix shown in Figure \ref{fig:corrmetr} highlights that certain parameters are more relevant for specific quantum devices. For the IBM Rochester device, the most important parameter for \textbf{gate overhead} is the \textbf{two-qubit gate percentage}, whereas for Aspen-16 is \textbf{the maximal degree}. The most important parameters for \textbf{fidelity decrease} of both devices are the \textbf{maximal and minimal degree of the qubit interaction graph, the number of qubits and gates} and  \textbf{ the two-qubit gate percentage}. In contrast, for the Surface-97 device, the most important parameters for \textbf{gate overhead} are the \textbf{number of qubits} and \textbf{the two-qubit gate percentage}, while the most important parameters for \textbf{fidelity decrease} are \textbf{the number of qubits} and \textbf{the maximal degree} of the qubit interaction graph. \textbf{Latency overhead} does not appear to be related to these structural parameters, so we will investigate this metric further in future work with other parameters. These observations suggest that:
\begin{itemize}
    \item \textbf{Interaction graph parameters} are more relevant for the mapping outcomes of Aspen-16 and IBM Rochester devices than for Surface-97. We can see that the majority of structural parameters are highly correlated with the circuit fidelity decrease. The main reasoning behind this is that these processors have much less connected coupling graphs; in other words, the sparser the coupling graph, the strongest the correlation with the interaction graph parameters.  In our case, Aspen-16 has the most restricted coupling graph connectivity, and consequently, its mapping metrics have the highest correlation with interaction graph properties.
    \item Two-qubit gate percentage, as expected, shows a very high correlation with the gate overhead metric regardless of the device. Other \textbf{size-related parameters} (number of qubits and gates) are highly correlated with the fidelity decrease of Aspen-16 and IBM Rochester devices due to again limited connectivity of their coupling graphs as well as smaller device size. On the other hand, the number of qubits only correlates with the gate overhead of Surface-97, which can be attributed to the fact it is a much larger device where we could run much bigger and more complex circuits that would then lead to inevitably long routing paths between at least some of the qubits.
\end{itemize}

iii) The two new topologies used for these experiments have quite similar structures (just in different scales of qubit range), and consequently, experiments showcased similar patterns. In future work, we plan to expand our analysis by including additional device topologies.

iv)  In cases where there is no correlation between interaction graph parameters and certain results (such as latency overhead and minimal degree), it suggests that other structural parameters may have played a more significant role. In our future work, we plan to investigate additional parameters such as gate-dependency critical paths and parallelism, which are discussed in Sec. \ref{Sec6}. Similar findings were also observed in a previous study \cite{tomesh2022supermarq}, demonstrating differences between topologies.

\begin{figure}[h!]
	\centering
	\includegraphics[width=0.9\linewidth]{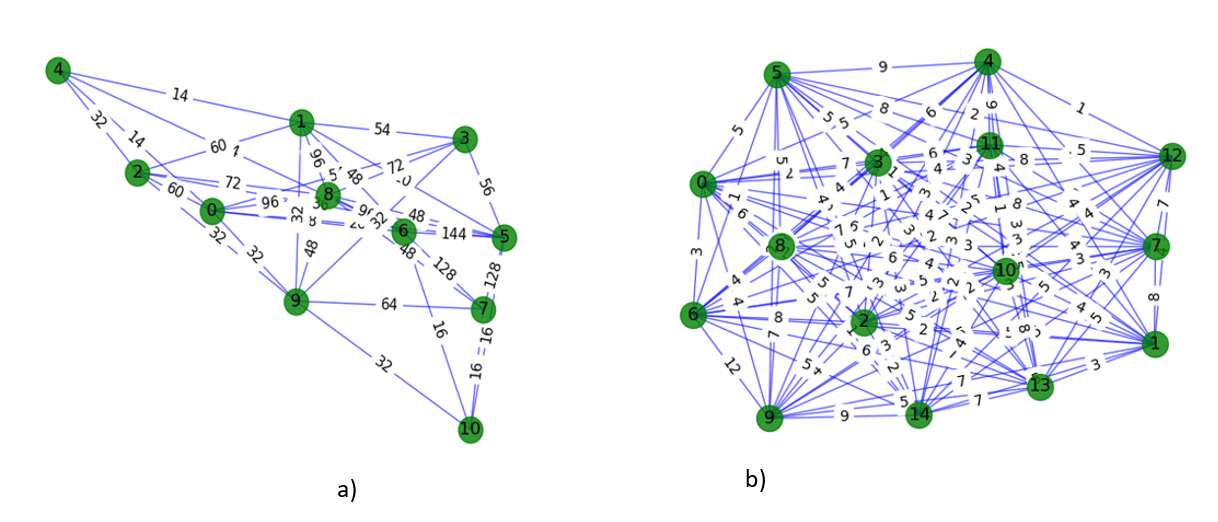}
	\caption{Qubit interaction graphs for circuits belonging to cluster 0 (a) and to cluster 3 (b).}
 	\label{fig:intergraphsstruct}
\end{figure}

To further investigate the benchmark cluster-device relationship, we continued by observing the circuits belonging to the same clusters. We noticed that (Fig. \ref{fig:intergraphsstruct}): cluster 0 consists of sparse, low-degree graphs and mostly RevLib circuits; cluster 1 is composed of circuits of a very large standard deviation of weight distribution; cluster 2 includes grid-like-shaped circuits with mostly QUEKO benchmarks; cluster 3 has the densest graphs with highest node degree, mostly consisting of randomly generated circuits; and finally, cluster 4 contains circuits with large average shortest path, mostly QUEKO circuits based on some existing algorithms \cite{Queko}.

As expected, the sparse graphs of low node degree in Cluster 0, which are easier to map to the 2D-grid-resembling qubit topology,  required the lowest amount of additional SWAPs, but due to specific, algorithm-based structure, could not be well optimized in terms of depth (more difficult to parallelize operations). Cluster 0 is the only cluster with circuits whose fidelity did not drop 100\% 

\begin{figure}[h!]
    \centering
    \centerline{
\subfigure[]{
    \includegraphics[width=0.5 \linewidth]{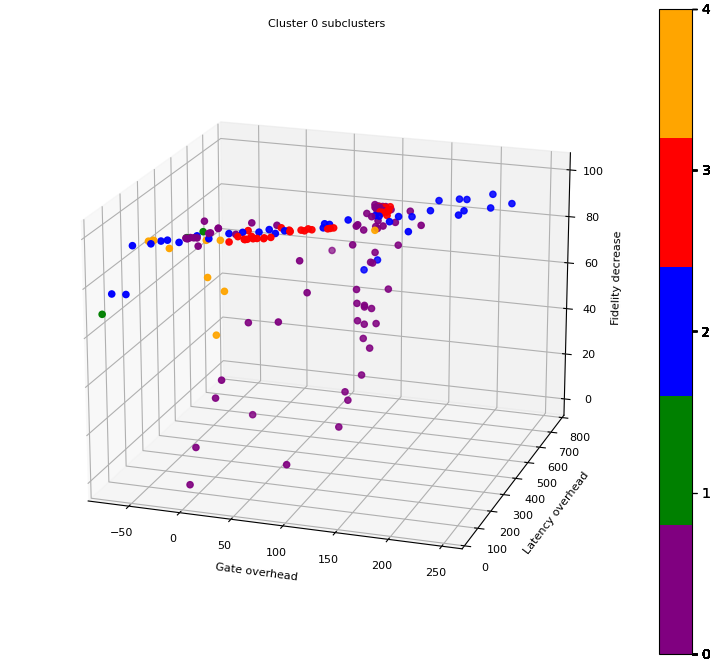}
    \label{fig:clusterresultsibm}}
    \subfigure[]{\includegraphics[width=0.5 \linewidth]{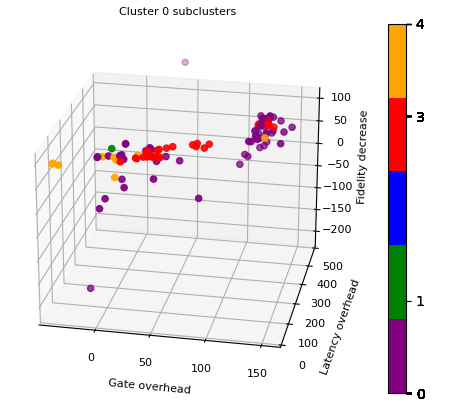}
    \label{fig:clusterresultsaspen}}}
    \caption{Quantum circuit mapping metrics vs. clusters of quantum circuits when targeting IBM Rochester (a) and Aspen-16 (b) topologies.}
    \label{fig:clusterresults2}
\end{figure}

On the other hand, the 2D-grid qubit topology, which is the most common state-of-the-art for quantum chips, could not handle well the dense graphs belonging to cluster 3, most of which are random circuits. However, they did perform fine in terms of their latency. What is also interesting, based on these outcomes, is that having, for instance, high average shortest path (like circuits in cluster 4), leads to low latency overhead - as explained in Sec. \ref{Sec5.1}, which means that the circuit depth was not extended so much.  That was as expected, considering that it means that those circuits are much less connected and easier to parallelize. 

Further, we have also analyzed the relationship between different circuit clusters and the mapping performance metrics for the experiments performed with the latter two quantum devices, the 53q Rochester and the 16q Aspen processor (see Fig. \ref{fig:clusterresults2}). This time we clearly see different outcomes. For instance, cluster 0 is not anymore outperforming the others in terms of gate overhead - cluster 4 shows the lowest gate overhead of  $\sim12\%$; cluster 3 fluctuates much more in terms of latency - it goes up to $\sim450\%$ instead of the previous $\sim150\%$ ; and cluster 4 is doing way better in terms of fidelity decrease- $\sim90\%$ instead of previous  $\sim100\%$. This is more evident for the Rochester device as the number of circuits included is significantly larger. As 16q-Aspen is on a smaller scale (lower number of qubits) similar to Rochester device in terms of connectivity, we also notice that they have similarly distributed clusters regarding mapping metrics. The data points in Fig. \ref{fig:clusterresultsaspen} could even be a subset of those in Fig. \ref{fig:clusterresultsibm}. This outcome means that other devices with similar topology and higher numbers of qubits would still show similar patterns.

We discuss other possible reasoning behind the results in \emph{Future work} section.

\section{Discussion and future work}
\label{Sec6}

In Sec. \ref{Sec3}, we mentioned that for completing the description of the structure of quantum circuits, in addition to the interaction graph, we also require gate dependency graph properties. Gate dependency graphs can give insight into how a circuit evolves in time. The critical path within the graph is the most relevant property as it is related to the parallelization degree of the gates, which directly influences the circuit depth. This would also help to explore the oracles or other patterns and repetitions within the circuit. In addition to gate dependency graphs, properties like the amount of parallelism in the circuit (gate density), measurement, and idle gates are influencing the success rate of the circuit a lot \cite{tomesh2022supermarq}.  

In addition to this, we must not underestimate the role of the mapping technique in these outcomes. For example, including features like look-ahead/back approaches or optimal initial qubit placement would probably have a stronger influence in terms of mapping results when used on circuits with already predefined, steady, and repetitive structures. To verify this assumption, we plan to compare the performance of quantum circuits when using different types of mappers and optimization properties to investigate the mapper-circuit relationship in contrast to the device-circuit relationship demonstrated in this paper. That could then lead to providing  guidelines for designing and optimizing algorithm-aware mapping techniques. To this purpose, structured design space exploration methodologies can be used as pointed out in  \cite{bandic2020structured}.

To conclude, in our future work, we would like to explore further: i) other structural parameters of quantum circuits based on gate dependency graphs such as critical path, the density of gates per layer, and amount of measurement and idle gates. With this, we will ensure to encapsulate all structural perspectives of quantum circuits when performing benchmark clustering and profiling; ii) how these observed patterns (with current parameters and additional ones) can help us to predict the mapping performance of new circuit samples assigned to our clusters, without actually running them on the device; iii) how exactly the interaction graph and coupling graph similarity relate to the mapping result; and iv) investigate a relationship between interaction graphs and gate dependencies with the chosen mapping technique and to which extend that affects the circuit mapping performance on-chip. For this, we will include more compiling options when performing comparisons. This insight into a circuit structure could help us compare and improve currently existing mapping techniques and enable us to have algorithm-driven mappers and quantum devices.

\section{Conclusion}
\label{Sec7}

Current quantum devices are still bounded by size and noise and can only handle small and simple quantum algorithms. To execute quantum algorithms, expressed as quantum circuits, on these error-prone and resource-constrained devices, they need to be adapted to overcome those limitations and therefore prevent additional errors. That process is referred to as the mapping of quantum circuits and represents a complex optimization problem that is dependent on both, processor and algorithm properties. In addition to hardware properties, in this paper, we have analyzed how the structure of quantum circuits affects their mapping performance. Our selected quantum circuits were characterized in terms of not only standard parameters, such as the number of qubits and gates and percentage of 2-qubit gates, but also in terms of their interaction graph (i.e., graph theory-based) parameters that include average shortest path, minimal and maximal node degree, and standard deviation of the edge-weight distribution. Our results show a strong correlation between these parameters and circuit mapping metrics: gate overhead, latency overhead, and fidelity decrease increased with the increase in all the chosen parameters. The effect of these parameters varies across different devices and metrics. For example, the degree parameter has a larger impact on fidelity decrease for the IBM Rochester device than for the Surface-97 device. From these findings, we can identify the preferred devices for an algorithm with specific individual metrics. Furthermore, after clustering the circuits based on mentioned parameters, we found patterns in mapping performance (in terms of the three mentioned metrics) of the circuits belonging to the same cluster, when mapped using the same technique on the same device. For instance, clusters with simpler, low node-degree graphs showed better performance when targeting a 2D-grid topology regarding gate overhead, whereas clusters consisting of complex and dense circuits outperformed others in latency. On the other hand, different performance results were noted when running the same groups of circuits on two other less-connected devices: size parameters like the number of qubits were far less relevant, and synthetic circuits outperformed real ones (which was not the case for Surface-97), and finally, the correlation between clusters of benchmarks and mapping results was unlike to the previously obtained ones. It was also shown that the way circuits were created is very related to their structure and impacts the results (e.g., if they were uniformly randomly generated circuits), as those circuits were in most cases grouped in the same clusters. Finally, we could see how the clusters scale with different mapping metrics. For instance, in one of the clusters, gate overhead scales linearly with latency overhead; in another, gate overhead is constantly within a specific range regardless of the increase in latency. 

The proposed method and current findings will help to enhance circuit mapping techniques  by including information about the structure of the circuit as well as to have a deeper understanding on the disparity of the observed outcomes when executing different quantum algorithms. In addition, structural parameters of circuits could be used to predict their fidelity decrease and gate and latency overhead for some specific processor and compilation technique without running them on actual devices.This could help to analyze and perform a design space exploration as well as codesign of current compilers, quantum processors and quantum applications. Ultimately, this process contributes to the development of application-specific quantum systems, where algorithms will be run with higher performances.

Quantum circuits are also used as benchmarks for evaluating mapping and quantum processors. However, the quantum community still does not agree on one benchmark set used, which resulted in an overwhelming amount of sources of quantum circuits. In this work, we have created a soon-to-be open-sourced easy-to-use benchmark collection having benchmarks from various sources cataloged in folders based on how they are implemented (e.g., based on a real algorithm, random, application-based), the language they are written in, and their size. The set also contains various scripts for translating circuits from one language to another, circuit interaction graphs, and profiling results, as described in this paper. We hope this collection will be useful for testing new quantum processors, updated regularly by the research community to keep up with the new technologies, compilers, programming languages, and most importantly applications, and eliminate the over-the-top amount of benchmark sources.

\section{Acknowledgements}

The authors sincerely appreciate the contribution of Nikiforos Paraskevopoulos in creating the benchmark collection and its documentation, as well as scientific discussions with Prof. Eduard Alarcon (UPC). MB and SF would also like to acknowledge funding from Intel Corporation. This work has been partially supported by the Spanish Ministerio de Ciencia e Innovación and European ERDF under grant PID2021-123627OB-C51 and by the QuantERA grant EQUIP, by the Ministerio de Ciencia e Innovaci\'{o}n and Agencia Estatal de Investigaci\'{o}n, MCIN/AEI/10.13039/501100011033, and by the European Union “NextGenerationEU”/PRTR” (CGA).

\section{Declarations}

\subsection{Competing interest}
Not applicable.

\subsection{Authors' contributions}

Author M.B. wrote the main manuscript which was reviewed and approved by authors C.G.A and S.F.

\subsection{Availability of data and materials}

 Once the final reviews are completed, the data will be accessible on \cite{qbench}.
\bibliography{mybibliography}

\newpage
\section{Appendices}
\label{App}

\begin{figure}[h!]
	\centering
	\includegraphics[width=\linewidth]{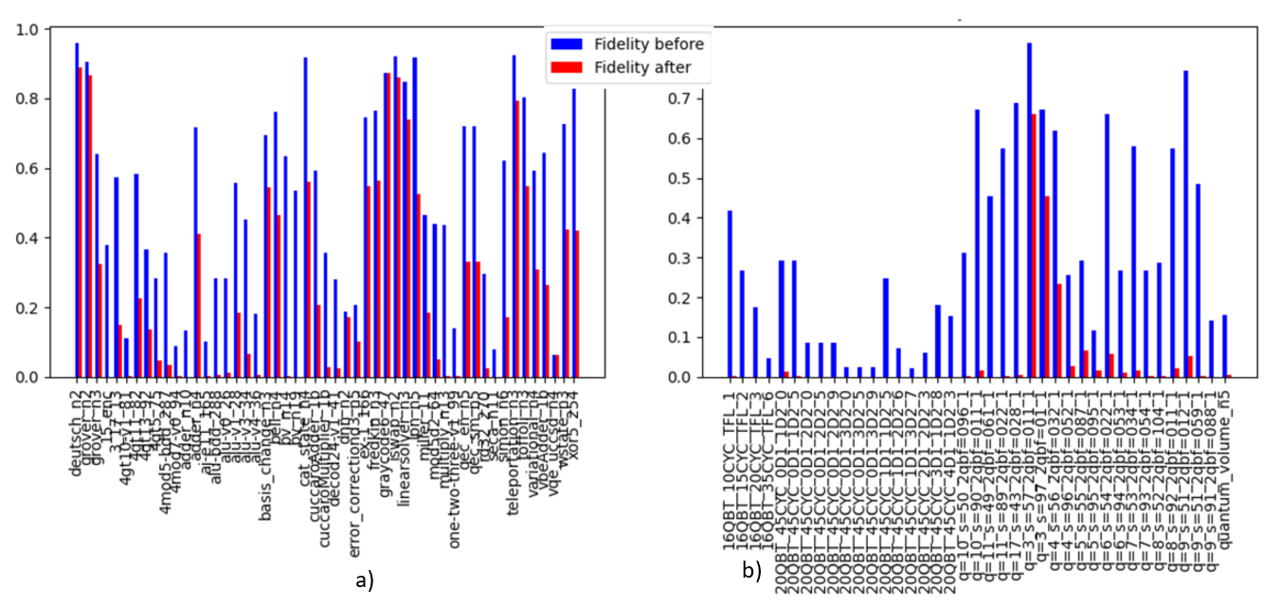
 }
	\caption{Fidelity decrease for real circuits (a) and synthetically generated ones (b). In this figure we included only the benchmarks whose fidelity was higher than 10\% to begin with.}
 	\label{fig:barchartfidelity}
\end{figure}

\begin{figure*}[h]
    \centering
    \centerline{
\subfigure[]{\includegraphics[width=0.46\linewidth]{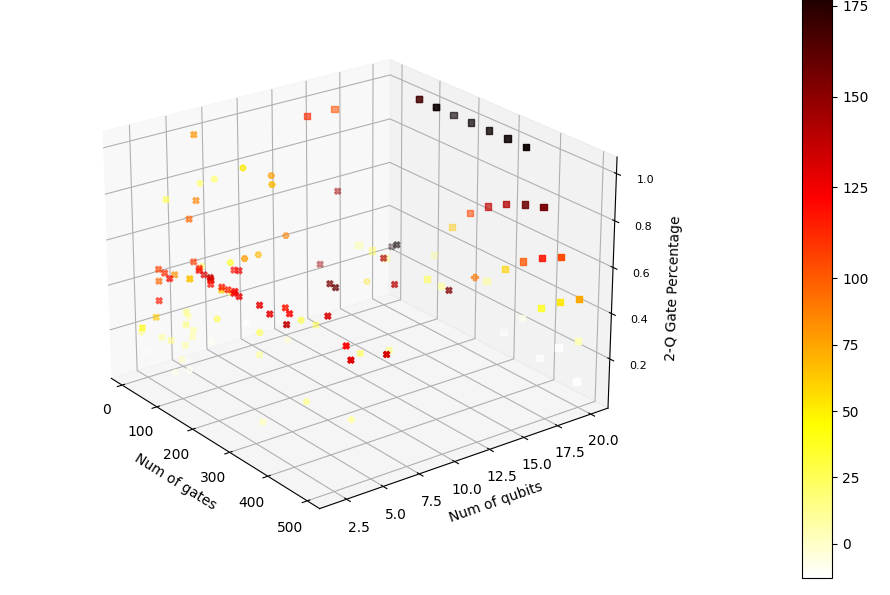}
\label{fig:ibm1}}
\subfigure[]{\includegraphics[width=0.41\linewidth]{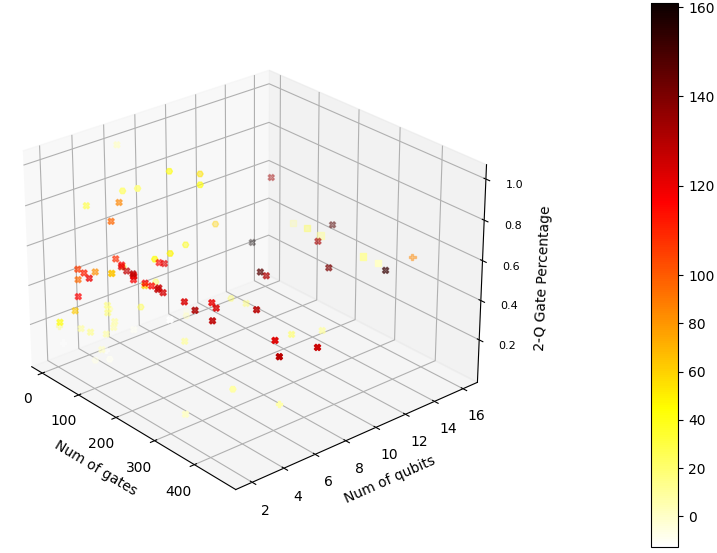}
\label{fig:aspen1}}}
\centerline{
\subfigure[]{\includegraphics[width=0.41\linewidth]{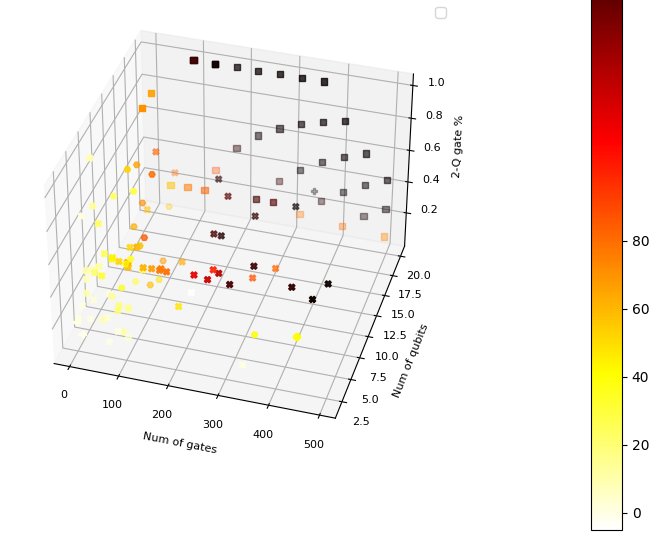}
\label{fig:ibm2}}
\subfigure[]{\includegraphics[width=0.41\linewidth]{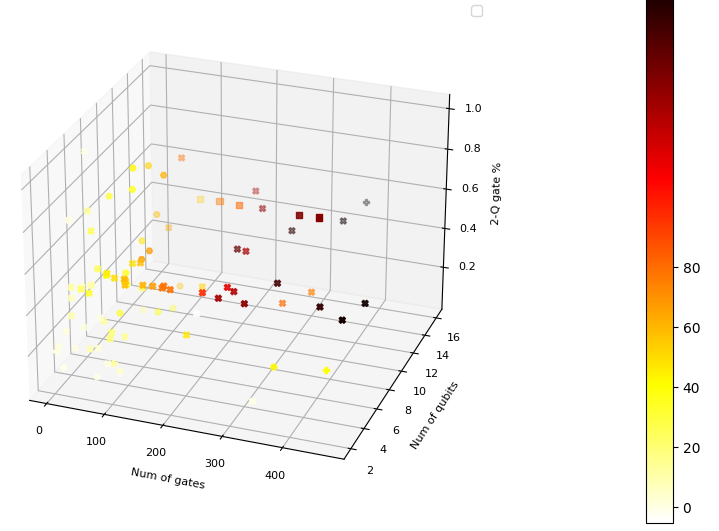} 
\label{fig:aspen2}}}
\centerline{
\subfigure[]{\includegraphics[width=0.4\linewidth]{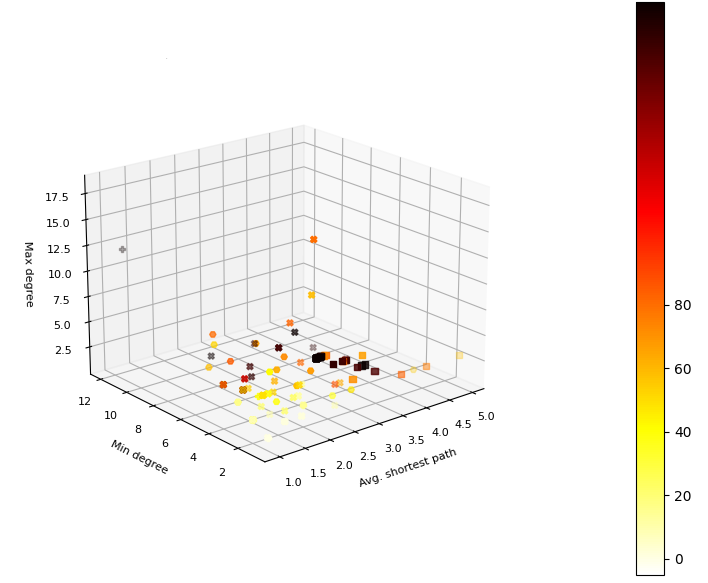}
\label{fig:ibm3}}
\subfigure[]{\includegraphics[width=0.46\linewidth]{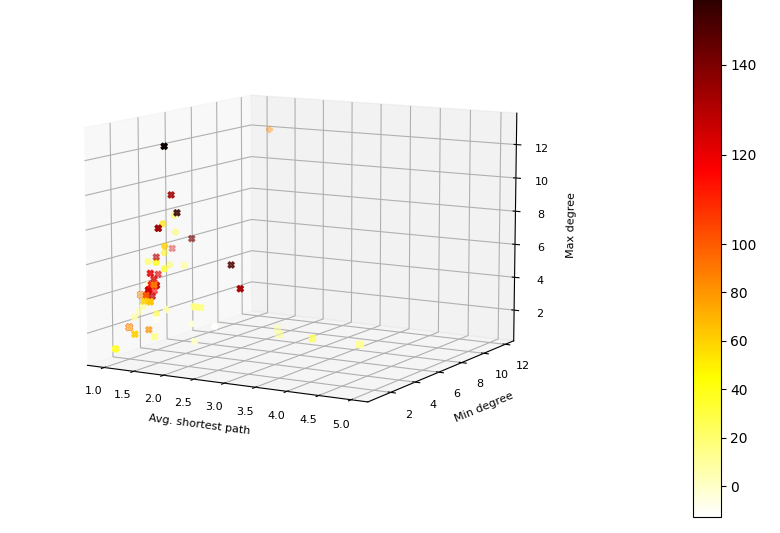} 
\label{fig:aspen3}}}
    \caption{Results of the circuit compilation when mapping different quantum  circuits (Random, QUEKO, Reversible arithmetic circuits, Quantum-algorithm based circuits) to the IBM Rochester (left) and Aspen 16q (right) device toplogies using the MinExtend mapper in terms of: a) and b) Gate overhead and size parameters; c) and d) Fidelity decrease and size parameters; e) Fidelity decrease and IG parameters and f) Gate overhead and IG parameters.}
\label{fig:ibm_aspen}
\end{figure*}

\newpage

\begin{table}[!t]
    \fontsize{5pt}{5pt}\selectfont
    \centering
    \caption{Mapping overhead results containing a sample of used quantum circuits with their properties for Surface-97 device}
    \begin{tabularx}{\textwidth}{|l|X|X|X|X|X|X|X|X|X|X|}
    \hline
        \textbf{Benchmark} & \textbf{Qubit count} & \textbf{Gate count} & \textbf{2q gate \%} & \textbf{Avg. hop count} & \textbf{Max degree} & \textbf{Min. degree} & \textbf{$\sigma$ adj. mat.} & \textbf{Gate inc.} & \textbf{Laten- cy inc.} & \textbf{Fide- lity dec.} \\ \hline 
        15\_enc & 15 & 125 & 0.27 & 1.73 & 10 & 2 & 0.00 & 820.00 & 583.33 & 99.99 \\ \hline
        16QBT15CYCTFL2 & 16 & 197 & 0.22 & 4.23 & 3 & 1 & 1.00 & 556.35 & 180.91 & 100.00 \\ \hline
        16QBT20CYCTFL3 & 16 & 261 & 0.22 & 3.70 & 3 & 1 & 1.00 & 318.77 & 106.85 & 99.97 \\ \hline
        16QBT35CYCTFL6 & 16 & 455 & 0.22 & 3.47 & 3 & 2 & 2.00 & 251.87 & 127.17 & 100.00 \\ \hline
        16QBT10CYCTFL1 & 16 & 131 & 0.22 & 4.97 & 3 & 1 & 0 & 474.81 & 216.22 & 99.77 \\ \hline
        20QBT45CYC0D11D20 & 19 & 135 & 0.33 & 2.88 & 6 & 1 & 1.00 & 208.89 & 317.39 & 95.09 \\ \hline
        20QBT45CYC0D11D25 & 13 & 135 & 0.33 & 2.69 & 4 & 1 & 1.00 & 343.70 & 300.00 & 99.05 \\ \hline
        20QBT45CYC0D12D20 & 20 & 270 & 0.33 & 2.47 & 6 & 1 & 1.00 & 714.81 & 658.24 & 100.00 \\ \hline
        20QBT45CYC0D12D25 & 20 & 270 & 0.33 & 2.54 & 6 & 1 & 1.00 & 581.48 & 448.35 & 100.00 \\ \hline
        20QBT45CYC0D12D29 & 20 & 270 & 0.33 & 2.78 & 5 & 1 & 1.00 & 507.78 & 384.62 & 100.00 \\ \hline
        20QBT45CYC0D13D20 & 20 & 405 & 0.33 & 2.30 & 6 & 2 & 1.41 & 539.26 & 384.56 & 100.00 \\ \hline
        20QBT45CYC1D12D26 & 20 & 360 & 0.25 & 2.54 & 6 & 1 & 1.00 & 435.00 & 164.09 & 100.00 \\ \hline
        20QBT45CYC1D13D27 & 20 & 495 & 0.27 & 2.34 & 6 & 2 & 1.41 & 472.53 & 249.12 & 100.00 \\ \hline
        20QBT45CYC2D12D23 & 20 & 450 & 0.20 & 2.55 & 6 & 1 & 1.00 & 281.33 & 76.38 & 100.00 \\ \hline
        20QBT45CYC4D11D23 & 20 & 495 & 0.09 & 2.95 & 5 & 1 & 0.00 & 134.34 & -27.59 & 99.98 \\ \hline
        3\_17\_13 & 3 & 102 & 0.17 & 1.00 & 2 & 2 & 2.83 & 128.43 & 478.38 & 74.26 \\ \hline
        4gt10-v1\_81 & 5 & 424 & 0.16 & 1.00 & 4 & 4 & 3.87 & 78.54 & 234.46 & 97.68 \\ \hline
        4gt11\_82 & 5 & 79 & 0.23 & 1.30 & 4 & 2 & 1.41 & 116.46 & 378.57 & 61.60 \\ \hline
        4gt13\_92 & 5 & 190 & 0.16 & 1.40 & 4 & 2 & 2.65 & 39.47 & 159.70 & 62.93 \\ \hline
        4gt5\_75 & 5 & 239 & 0.16 & 1.10 & 4 & 3 & 2.45 & 64.02 & 260.24 & 83.34 \\ \hline
        4mod5-bdd\_287 & 7 & 196 & 0.16 & 1.62 & 5 & 1 & 1.73 & 112.76 & 307.14 & 90.71 \\ \hline
        4mod7-v0\_94 & 5 & 466 & 0.15 & 1.10 & 4 & 3 & 4.47 & 73.39 & 236.20 & 98.00 \\ \hline
        adder\_n10 & 10 & 328 & 0.20 & 2.33 & 4 & 1 & 1.41 & 229.57 & 1590.32 & 99.92 \\ \hline
        adder\_n4 & 4 & 63 & 0.16 & 1.33 & 2 & 2 & 1.41 & 84.13 & 195.65 & 42.71 \\ \hline
        aj-e11\_165 & 5 & 433 & 0.16 & 1.00 & 4 & 4 & 4.36 & 76.91 & 250.99 & 97.68 \\ \hline
        alu-bdd\_288 & 7 & 240 & 0.16 & 1.43 & 5 & 2 & 1.73 & 153.75 & 336.90 & 97.68 \\ \hline
        alu-v0\_26 & 5 & 240 & 0.16 & 1.20 & 4 & 3 & 3.16 & 121.25 & 341.67 & 95.48 \\ \hline
        alu-v1\_28 & 5 & 105 & 0.17 & 1.30 & 4 & 2 & 1.41 & 98.10 & 224.32 & 67.00 \\ \hline
        alu-v3\_34 & 5 & 148 & 0.16 & 1.30 & 4 & 2 & 1.73 & 124.32 & 283.02 & 85.25 \\ \hline
        alu-v4\_36 & 5 & 329 & 0.16 & 1.10 & 4 & 3 & 3.32 & 90.88 & 222.41 & 96.34 \\ \hline
        basis\_change\_n3 & 3 & 79 & 0.13 & 1.33 & 2 & 1 & 2.65 & 34.18 & 74.07 & 21.84 \\ \hline
        bell\_n4 & 4 & 66 & 0.11 & 1.67 & 2 & 1 & 1.00 & 81.82 & 62.75 & 38.92 \\ \hline
        bv\_n14 & 14 & 94 & 0.14 & 1.86 & 13 & 1 & 0.00 & 593.62 & 285.37 & 99.50 \\ \hline
        cat\_state\_n4 & 4 & 11 & 0.27 & 1.67 & 2 & 1 & 0.00 & 490.91 & 780.00 & 38.92 \\ \hline
        cuccaroAdder\_1b & 4 & 83 & 0.20 & 1.33 & 3 & 1 & 1.41 & 130.12 & 2025.00 & 65.07 \\ \hline
        cuccaroMultiplier\_1b & 6 & 180 & 0.18 & 1.20 & 4 & 2 & 1.00 & 151.11 & 958.82 & 92.57 \\ \hline
        decod24-v1\_41 & 5 & 241 & 0.16 & 1.20 & 4 & 2 & 2.65 & 92.95 & 287.21 & 91.35 \\ \hline
        deutsch\_n2 & 2 & 10 & 0.10 & 1.00 & 1 & 1 & 0.00 & 60.00 & 180.00 & 7.39 \\ \hline
        dnn\_n2 & 2 & 422 & 0.10 & 1.00 & 1 & 1 & 24.25 & 2.13 & 85.00 & 7.89 \\ \hline
        error\_correctiond3\_n5 & 5 & 278 & 0.18 & 1.50 & 4 & 1 & 5.92 & 0.36 & 64.91 & 51.67 \\ \hline
        ex-1\_166 & 3 & 53 & 0.17 & 1.00 & 2 & 2 & 1.41 & 45.28 & 210.53 & 26.43 \\ \hline
        fredkin\_n3 & 3 & 51 & 0.16 & 1.00 & 2 & 2 & 1.41 & 43.14 & 285.00 & 26.17 \\ \hline
        graycode6\_47 & 6 & 15 & 0.33 & 2.33 & 2 & 1 & 0.00 & 0.00 & 166.67 & 0.00 \\ \hline
        grover\_n2 & 2 & 30 & 0.07 & 1.00 & 1 & 1 & 1.00 & -36.67 & 6.25 & 4.51 \\ \hline
        grover\_n3 & 3 & 102 & 0.12 & 1.00 & 2 & 2 & 1.00 & 45.10 & 1160.00 & 49.15 \\ \hline
        iswap\_n2 & 2 & 21 & 0.10 & 1.00 & 1 & 1 & 1.00 & 4.76 & 111.11 & 6.55 \\ \hline
        linearsolver\_n3 & 3 & 43 & 0.09 & 1.33 & 2 & 1 & 1.00 & 4.65 & 108.70 & 12.67 \\ \hline
        lpn\_n5 & 5 & 24 & 0.08 & 1.33 & 2 & 1 & 0.00 & 216.67 & 272.73 & 42.61 \\ \hline
        miller\_11 & 3 & 144 & 0.16 & 1.00 & 2 & 2 & 3.87 & 52.08 & 327.45 & 60.40 \\ \hline
        mod5d2\_64 & 5 & 151 & 0.17 & 1.20 & 4 & 2 & 1.73 & 135.76 & 358.49 & 88.35 \\ \hline
        multiply\_n13 & 11 & 144 & 0.18 & 2.44 & 4 & 1 & 0.00 & 393.75 & 2311.11 & 99.47 \\ \hline
        one-two-three-v1\_99 & 5 & 378 & 0.16 & 1.10 & 4 & 3 & 3.61 & 87.83 & 222.56 & 97.52 \\ \hline
        q=10\_s=50\_2qbf=096\_1 & 10 & 60 & 0.80 & 1.36 & 8 & 3 & 1.00 & 1125.00 & 326.23 & 99.79 \\ \hline
        q=10\_s=90\_2qbf=011\_1 & 10 & 100 & 0.10 & 2.67 & 3 & 1 & 0.00 & 393.00 & 18.00 & 97.78 \\ \hline
        q=11\_s=49\_2qbf=061\_1 & 11 & 60 & 0.52 & 1.62 & 7 & 2 & 0.00 & 1466.67 & 368.85 & 99.97 \\ \hline
        q=11\_s=89\_2qbf=022\_1 & 11 & 100 & 0.17 & 1.84 & 6 & 1 & 0.00 & 567.00 & 99.01 & 99.47 \\ \hline
        q=17\_s=43\_2qbf=028\_1 & 17 & 60 & 0.20 & 3.97 & 3 & 1 & 0.00 & 896.67 & 120.00 & 99.27 \\ \hline
        q=3\_s=57\_2qbf=011\_1 & 3 & 60 & 0.05 & 1.33 & 2 & 1 & 0.00 & 38.33 & -20.00 & 21.28 \\ \hline
        q=3\_s=97\_2qbf=01\_1 & 3 & 100 & 0.10 & 1.00 & 2 & 2 & 1.41 & 32.00 & -10.00 & 32.11 \\ \hline
        q=4\_s=56\_2qbf=032\_1 & 4 & 60 & 0.28 & 1.00 & 3 & 3 & 1.00 & 171.67 & 126.67 & 62.35 \\ \hline
        q=4\_s=96\_2qbf=052\_1 & 4 & 100 & 0.54 & 1.00 & 3 & 3 & 5.20 & 248.00 & 156.44 & 89.90 \\ \hline
        q=5\_s=55\_2qbf=087\_1 & 5 & 60 & 0.85 & 1.00 & 4 & 4 & 3.16 & 268.33 & 191.80 & 77.17 \\ \hline
        q=5\_s=95\_2qbf=095\_1 & 5 & 100 & 0.90 & 1.00 & 4 & 4 & 4.36 & 225.00 & 182.00 & 87.18 \\ \hline
        q=6\_s=54\_2qbf=022\_1 & 6 & 60 & 0.23 & 1.47 & 3 & 2 & 0.00 & 443.33 & 125.00 & 91.43 \\ \hline
        q=6s=942qbf=053\_1 & 6 & 100 & 0.52 & 1.00 & 5 & 5 & 2.24 & 354.00 & 190.10 & 96.22 \\ \hline
        q=7s=532qbf=0341 & 7 & 60 & 0.33 & 1.33 & 6 & 2 & 0.00 & 660.00 & 200.00 & 97.48 \\ \hline
        q=7\_s=93\_2qbf=054\_1 & 7 & 100 & 0.52 & 1.14 & 6 & 4 & 1.73 & 522.00 & 214.85 & 99.15 \\ \hline
        q=8\_s=52\_2qbf=104\_1 & 8 & 60 & 0.87 & 1.11 & 7 & 5 & 1.00 & 825.00 & 291.80 & 98.91 \\ \hline
        q=8\_s=92\_2qbf=011\_1 & 8 & 100 & 0.17 & 1.57 & 5 & 1 & 0.00 & 351.00 & 48.00 & 96.44 \\ \hline
        q=9\_s=51\_2qbf=012\_1 & 9 & 60 & 0.12 & 2.29 & 3 & 1 & 0.00 & 480.00 & 75.00 & 93.24 \\ \hline
        q=9\_s=51\_2qbf=059\_1 & 9 & 60 & 0.47 & 1.47 & 6 & 2 & 0.00 & 943.33 & 253.33 & 99.43 \\ \hline
        q=9\_s=91\_2qbf=088\_1 & 9 & 100 & 0.81 & 1.08 & 8 & 5 & 1.41 & 639.00 & 259.41 & 99.71 \\ \hline
        qec\_en\_n5 & 5 & 61 & 0.16 & 1.60 & 4 & 1 & 1.00 & 106.56 & 138.46 & 54.00 \\ \hline
        qec\_sm\_n5 & 5 & 61 & 0.16 & 1.60 & 4 & 1 & 1.00 & 106.56 & 138.46 & 54.00 \\ \hline
        quantum\_volume\_n5 & 5 & 411 & 0.12 & 1.20 & 4 & 2 & 6.93 & 101.46 & 114.41 & 97.23 \\ \hline
        rd32\_270 & 5 & 236 & 0.15 & 1.10 & 4 & 3 & 2.45 & 98.73 & 252.94 & 92.03 \\ \hline
        seca\_n11 & 11 & 396 & 0.21 & 1.95 & 6 & 2 & 1.73 & 228.28 & 536.23 & 99.99 \\ \hline
        simon\_n6 & 5 & 92 & 0.15 & 1.70 & 3 & 1 & 0.00 & 138.04 & 900.00 & 72.30 \\ \hline
        teleportation\_n3 & 3 & 20 & 0.10 & 1.33 & 2 & 1 & 0.00 & 55.00 & 125.00 & 14.07 \\ \hline
        toffoli\_n3 & 3 & 48 & 0.13 & 1.00 & 2 & 2 & 1.00 & 62.50 & 266.67 & 31.87 \\ \hline
        variational\_n4 & 4 & 94 & 0.17 & 1.67 & 2 & 1 & 2.83 & 72.34 & 150.00 & 47.80 \\ \hline
        vbeAdder\_1b & 4 & 74 & 0.19 & 1.17 & 3 & 2 & 1.00 & 122.97 & 2300.00 & 58.90 \\ \hline
        vqe\_uccsd\_n4 & 4 & 452 & 0.19 & 1.67 & 2 & 1 & 14.70 & -29.87 & 74.55 & 2.03 \\ \hline
        wstate\_n3 & 3 & 68 & 0.13 & 1.00 & 2 & 2 & 1.41 & 66.18 & 470.59 & 41.88 \\ \hline
        xor5\_254 & 6 & 17 & 0.29 & 1.67 & 5 & 1 & 0.00 & 429.41 & 537.50 & 51.57  \\ \hline
    \end{tabularx}
\label{tab:results_surface97}
\end{table}

\newpage
\begin{table}[!ht]
    \fontsize{5pt}{5pt}\selectfont
    \centering
    \caption{Mapping overhead results containing a sample of used quantum circuits with their properties for Aspen-16 device} 
    \begin{tabularx}{\textwidth}{|l|X|X|X|X|X|X|X|X|X|X|}
    \hline
        \textbf{Benchmark} & \textbf{Qubit count} & \textbf{Gate count} & \textbf{2q gate \%} & \textbf{Avg. hop count} & \textbf{Max degree} & \textbf{Min. degree} & \textbf{$\sigma$ adj. mat.} & \textbf{Gate inc.} & \textbf{Laten- cy inc.} & \textbf{Fide- lity dec.} \\ \hline 
        15\_enc & 15 & 53 & 0.64  & 1.73  & 10 & 2 & 0.00  & 133.96  & 344.64  & 79.82  \\ \hline
        16QBT\_10CYC\_TFL\_1 & 16 & 73 & 0.40  & 4.97  & 3 & 1 & 0.00  & 10.96  & 78.95  & 55.32  \\ \hline
        16QBT\_15CYC\_TFL\_2 & 16 & 109 & 0.40  & 4.23  & 3 & 1 & 1.00  & 24.77  & 151.79  & 77.17  \\ \hline
        16QBT\_20CYC\_TFL\_3 & 16 & 145 & 0.40  & 3.70  & 3 & 1 & 1.00  & 9.66  & 104.73  & 81.65  \\ \hline
        16QBT\_35CYC\_TFL\_6 & 16 & 253 & 0.40  & 3.47  & 3 & 2 & 2.00  & 9.88  & 84.38  & 94.39  \\ \hline
        3\_17\_13 & 3 & 36 & 0.47  & 1.00  & 2 & 2 & 2.83  & 111.11  & 341.03  & 21.96  \\ \hline
        4gt10-v1\_81 & 5 & 148 & 0.45  & 1.00  & 4 & 4 & 3.87  & 127.70  & 384.00  & 77.81  \\ \hline
        4gt11\_82 & 5 & 27 & 0.67  & 1.30  & 4 & 2 & 1.41  & 96.30  & 356.67  & 23.40  \\ \hline
        4gt13\_92 & 5 & 66 & 0.45  & 1.40  & 4 & 2 & 2.65  & 127.27  & 360.87  & 49.28  \\ \hline
        4gt5\_75 & 5 & 83 & 0.46  & 1.10  & 4 & 3 & 2.45  & 130.12  & 388.24  & 59.26  \\ \hline
        4mod5-bdd\_287 & 7 & 70 & 0.44  & 1.62  & 5 & 1 & 1.73  & 117.14  & 373.61  & 49.10  \\ \hline
        4mod7-v0\_94 & 5 & 162 & 0.44  & 1.10  & 4 & 3 & 4.47  & 124.07  & 370.30  & 78.28  \\ \hline
        adder\_n10 & 10 & 294 & 0.22  & 2.33  & 4 & 1 & 1.41  & 17.35  & 1948.48  & 80.84  \\ \hline
        adder\_n4 & 4 & 23 & 0.43  & 1.33  & 2 & 2 & 1.41  & 82.61  & 124.00  & 7.52  \\ \hline
        aj-e11\_165 & 5 & 151 & 0.46  & 1.00  & 4 & 4 & 4.36  & 126.49  & 383.55  & 78.84  \\ \hline
        alu-bdd\_288 & 7 & 84 & 0.45  & 1.43  & 5 & 2 & 1.73  & 121.43  & 354.12  & 58.81  \\ \hline
        alu-v0\_26 & 5 & 84 & 0.45  & 1.20  & 4 & 3 & 3.16  & 115.48  & 376.47  & 53.61  \\ \hline
        alu-v1\_28 & 5 & 37 & 0.49  & 1.30  & 4 & 2 & 1.41  & 108.11  & 342.11  & 25.31  \\ \hline
        alu-v3\_34 & 5 & 52 & 0.46  & 1.30  & 4 & 2 & 1.73  & 125.00  & 358.18  & 41.42  \\ \hline
        alu-v4\_36 & 5 & 115 & 0.44  & 1.10  & 4 & 3 & 3.32  & 117.39  & 358.47  & 62.04  \\ \hline
        basis\_change\_n3 & 3 & 79 & 0.13  & 1.33  & 2 & 1 & 2.65  & 0.00  & 46.91  & 0.00  \\ \hline
        bell\_n4 & 4 & 49 & 0.14  & 1.67  & 2 & 1 & 1.00  & 8.16  & 49.02  & 2.88  \\ \hline
        bv\_n14 & 14 & 41 & 0.32  & 1.86  & 13 & 1 & 0.00  & 160.98  & 266.67  & 63.94  \\ \hline
        cat\_state\_n4 & 4 & 4 & 0.75  & 1.67  & 2 & 1 & 0.00  & 25.00  & 100.00  & 0.18  \\ \hline
        cuccaroAdder\_1b & 4 & 73 & 0.23  & 1.33  & 3 & 1 & 1.41  & 6.85  & 1540.00  & 20.45  \\ \hline
        cuccaroMultiplier\_1b & 6 & 176 & 0.18  & 1.20  & 4 & 2 & 1.00  & 8.52  & 852.94  & 47.40  \\ \hline
        decod24-v1\_41 & 5 & 85 & 0.45  & 1.20  & 4 & 2 & 2.65  & 124.71  & 395.45  & 57.27  \\ \hline
        deutsch\_n2 & 2 & 5 & 0.20  & 1.00  & 1 & 1 & 0.00  & 20.00  & 14.29  & 0.18  \\ \hline
        dnn\_n2 & 2 & 338 & 0.12  & 1.00  & 1 & 1 & 24.25  & 0.00  & 97.65  & 0.00  \\ \hline
        error\_correctiond3\_n5 & 5 & 114 & 0.43  & 1.50  & 4 & 1 & 5.92  & 60.53  & 139.13  & 24.28  \\ \hline
        ex-1\_166 & 3 & 19 & 0.47  & 1.00  & 2 & 2 & 1.41  & 105.26  & 319.05  & 11.66  \\ \hline
        fredkin\_n3 & 3 & 19 & 0.42  & 1.00  & 2 & 2 & 1.41  & 94.74  & 209.09  & 7.35  \\ \hline
        graycode6\_47 & 6 & 5 & 1.00  & 2.33  & 2 & 1 & 0.00  & 0.00  & 150.00  & 0.00  \\ \hline
        grover\_n2 & 2 & 16 & 0.13  & 1.00  & 1 & 1 & 1.00  & -43.75  & -23.53  & -1.27  \\ \hline
        grover\_n3 & 3 & 89 & 0.13  & 1.00  & 2 & 2 & 1.00  & -19.10  & 945.45  & 11.59  \\ \hline
        iswap\_n2 & 2 & 9 & 0.22  & 1.00  & 1 & 1 & 1.00  & 44.44  & 27.27  & 0.72  \\ \hline
        linearsolver\_n3 & 3 & 23 & 0.17  & 1.33  & 2 & 1 & 1.00  & 8.70  & 56.00  & 0.36  \\ \hline
        lpn\_n5 & 5 & 11 & 0.18  & 1.33  & 2 & 1 & 0.00  & 18.18  & 50.00  & 2.53  \\ \hline
        miller\_11 & 3 & 50 & 0.46  & 1.00  & 2 & 2 & 3.87  & 120.00  & 373.58  & 31.05  \\ \hline
        mod5d2\_64 & 5 & 53 & 0.47  & 1.20  & 4 & 2 & 1.73  & 116.98  & 383.64  & 44.86  \\ \hline
        multiply\_n13 & 11 & 140 & 0.19  & 2.44  & 4 & 1 & 0.00  & 15.71  & 1263.64  & 51.37  \\ \hline
        one-two-three-v1\_99 & 5 & 132 & 0.45  & 1.10  & 4 & 3 & 3.61  & 115.15  & 371.11  & 67.01  \\ \hline
        q=10\_s=50\_2qbf=096\_1 & 10 & 60 & 0.80  & 1.36  & 8 & 3 & 1.00  & 53.33  & 159.02  & 53.28  \\ \hline
        q=10\_s=90\_2qbf=011\_1 & 10 & 100 & 0.10  & 2.67  & 3 & 1 & 0.00  & -11.00  & -2.00  & 32.82  \\ \hline
        q=11\_s=49\_2qbf=061\_1 & 11 & 60 & 0.52  & 1.62  & 7 & 2 & 0.00  & 60.00  & 186.89  & 59.35  \\ \hline
        q=11\_s=89\_2qbf=022\_1 & 11 & 100 & 0.17  & 1.84  & 6 & 1 & 0.00  & 28.00  & 47.52  & 57.84  \\ \hline
        q=3\_s=97\_2qbf=01\_1 & 3 & 100 & 0.10  & 1.00  & 2 & 2 & 1.41  & -10.00  & -5.00  & 4.68  \\ \hline
        q=4\_s=56\_2qbf=032\_1 & 4 & 60 & 0.28  & 1.00  & 3 & 3 & 1.00  & 3.33  & 65.00  & 14.57  \\ \hline
        q=4\_s=96\_2qbf=052\_1 & 4 & 100 & 0.54  & 1.00  & 3 & 3 & 5.20  & 12.00  & 133.66  & 31.15  \\ \hline
        q=5\_s=55\_2qbf=087\_1 & 5 & 60 & 0.85  & 1.00  & 4 & 4 & 3.16  & 18.33  & 132.79  & 24.69  \\ \hline
        q=5\_s=95\_2qbf=095\_1 & 5 & 100 & 0.90  & 1.00  & 4 & 4 & 4.36  & 15.00  & 97.00  & 30.00  \\ \hline
        q=6\_s=54\_2qbf=022\_1 & 6 & 60 & 0.23  & 1.47  & 3 & 2 & 0.00  & 25.00  & 63.33  & 35.89  \\ \hline
        q=6\_s=94\_2qbf=053\_1 & 6 & 100 & 0.52  & 1.00  & 5 & 5 & 2.24  & 42.00  & 151.49  & 67.72  \\ \hline
        q=7\_s=53\_2qbf=034\_1 & 7 & 60 & 0.33  & 1.33  & 6 & 2 & 0.00  & 20.00  & 71.67  & 38.32  \\ \hline
        q=7\_s=93\_2qbf=054\_1 & 7 & 100 & 0.52  & 1.14  & 6 & 4 & 1.73  & 48.00  & 181.19  & 68.06  \\ \hline
        q=8\_s=52\_2qbf=104\_1 & 8 & 60 & 0.87  & 1.11  & 7 & 5 & 1.00  & 38.33  & 131.15  & 42.13  \\ \hline
        q=8\_s=92\_2qbf=011\_1 & 8 & 100 & 0.17  & 1.57  & 5 & 1 & 0.00  & 1.00  & 17.00  & 34.26  \\ \hline
        q=9\_s=51\_2qbf=012\_1 & 9 & 60 & 0.12  & 2.29  & 3 & 1 & 0.00  & -5.00  & -40.00  & 11.88  \\ \hline
        q=9\_s=51\_2qbf=059\_1 & 9 & 60 & 0.47  & 1.47  & 6 & 2 & 0.00  & 43.33  & 106.67  & 47.29  \\ \hline
        q=9\_s=91\_2qbf=088\_1 & 9 & 100 & 0.81  & 1.08  & 8 & 5 & 1.41  & 40.00  & 120.79  & 61.37  \\ \hline
        qaoa\_n16 & 16 & 372 & 0.52  & 1.20  & 12 & 12 & 0.00  & 84.68  & 159.79  & 99.84  \\ \hline
        qaoa\_n6 & 6 & 414 & 0.13  & 1.40  & 3 & 3 & 3.00  & 6.28  & 49.28  & 38.52  \\ \hline
        qec\_en\_n5 & 5 & 25 & 0.40  & 1.60  & 4 & 1 & 1.00  & 64.00  & 167.86  & 11.02  \\ \hline
        qec\_sm\_n5 & 5 & 25 & 0.40  & 1.60  & 4 & 1 & 1.00  & 64.00  & 167.86  & 11.02  \\ \hline
        qft\_n15 & 15 & 540 & 0.39  & 1.00  & 14 & 14 & 0.00  & 35.74  & 123.11  & 98.59  \\ \hline
        quantum\_volume\_n5 & 5 & 338 & 0.15  & 1.20  & 4 & 2 & 6.93  & 7.10  & 66.76  & 42.23  \\ \hline
        rd32\_270 & 5 & 84 & 0.43  & 1.10  & 4 & 3 & 2.45  & 117.86  & 385.06  & 57.67  \\ \hline
        seca\_n11 & 11 & 333 & 0.25  & 1.95  & 6 & 2 & 1.73  & 7.81  & 510.00  & 75.03  \\ \hline
        shor\_15 & 11 & 4792 & 0.37  & 1.38  & 8 & 4 & 36.61  & 122.58  & 215.33  & 100.00  \\ \hline
        shor\_35 & 15 & 16529 & 0.37  & 1.41  & 12 & 5 & 79.04  & 120.47  & 216.06  & 100.00  \\ \hline
        simon\_n6 & 5 & 82 & 0.17  & 1.70  & 3 & 1 & 0.00  & -9.76  & 635.29  & 14.91  \\ \hline
        teleportation\_n3 & 3 & 8 & 0.25  & 1.33  & 2 & 1 & 0.00  & 62.50  & 50.00  & 0.90  \\ \hline
        toffoli\_n3 & 3 & 18 & 0.33  & 1.00  & 2 & 2 & 1.00  & 116.67  & 245.00  & 9.86  \\ \hline
        variational\_n4 & 4 & 54 & 0.30  & 1.67  & 2 & 1 & 2.83  & 14.81  & 103.57  & 1.43  \\ \hline
        vbeAdder\_1b & 4 & 70 & 0.20  & 1.17  & 3 & 2 & 1.00  & 5.71  & 1928.57  & 20.31  \\ \hline
        vqe\_uccsd\_n4 & 4 & 220 & 0.40  & 1.67  & 2 & 1 & 14.70  & -12.73  & 110.86  & -5.17  \\ \hline
        wstate\_n3 & 3 & 49 & 0.18  & 1.00  & 2 & 2 & 1.41  & 12.24  & 363.16  & 9.40  \\ \hline
        xor5\_254 & 6 & 7 & 0.71  & 1.67  & 5 & 1 & 0.00  & 85.71 & 400.00  & 13.30  \\ \hline
    \end{tabularx}
\label{tab:results_aspen-16}
\end{table}

\begin{table}[!ht]
    \fontsize{5pt}{5pt}\selectfont
    \centering
    \caption{Mapping overhead-based results containing a sample of used quantum circuits with their properties for IBM Rochester device}
    \begin{tabularx}{\textwidth}{|l|X|X|X|X|X|X|X|X|X|X|}
    \hline
        \textbf{Benchmark} & \textbf{Qubit count} & \textbf{Gate count} & \textbf{2q gate \%} & \textbf{Avg. hop count} & \textbf{Max degree} & \textbf{Min. degree} & \textbf{$\sigma$ adj. mat.} & \textbf{Gate inc.} & \textbf{Laten- cy inc.} & \textbf{Fide- lity dec.} \\ \hline 
                15\_enc & 15.00 & 53.00 & 0.64 & 1.73 & 10.00 & 2.00 & 0.00 & 145.28 & 448.21 & 82.50  \\ \hline
        16QBT\_10CYC\_TFL\_1 & 16.00 & 73.00 & 0.40 & 4.97 & 3.00 & 1.00 & 0.00 & 13.70 & 132.89 & 57.40  \\ \hline
        16QBT\_15CYC\_TFL\_2 & 16.00 & 109.00 & 0.40 & 4.23 & 3.00 & 1.00 & 1.00 & 22.94 & 127.68 & 76.06  \\ \hline
        16QBT\_20CYC\_TFL\_3 & 16.00 & 145.00 & 0.40 & 3.70 & 3.00 & 1.00 & 1.00 & 11.72 & 97.30 & 82.91  \\ \hline
        16QBT\_35CYC\_TFL\_6 & 16.00 & 253.00 & 0.40 & 3.47 & 3.00 & 2.00 & 2.00 & 20.95 & 119.14 & 97.12  \\ \hline
        20QBT\_45CYC\_0D1\_1D2\_0 & 15.00 & 45.00 & 1.00 & 2.88 & 6.00 & 1.00 & 1.00 & 102.22 & 516.67 & 66.51  \\ \hline
        20QBT\_45CYC\_0D1\_1D2\_5 & 13.00 & 45.00 & 1.00 & 2.69 & 4.00 & 1.00 & 1.00 & 113.33 & 533.33 & 70.26  \\ \hline
        20QBT\_45CYC\_0D1\_2D2\_0 & 20.00 & 90.00 & 1.00 & 2.47 & 6.00 & 1.00 & 1.00 & 173.33 & 536.56 & 97.55  \\ \hline
        20QBT\_45CYC\_0D1\_2D2\_5 & 20.00 & 90.00 & 1.00 & 2.54 & 6.00 & 1.00 & 1.00 & 175.56 & 293.55 & 97.67  \\ \hline
        20QBT\_45CYC\_0D1\_2D2\_9 & 20.00 & 90.00 & 1.00 & 2.78 & 5.00 & 1.00 & 1.00 & 163.33 & 411.83 & 96.97  \\ \hline
        20QBT\_45CYC\_0D1\_3D2\_0 & 20.00 & 135.00 & 1.00 & 2.30 & 6.00 & 2.00 & 1.41 & 202.22 & 523.19 & 99.85  \\ \hline
        20QBT\_45CYC\_0D1\_3D2\_5 & 20.00 & 135.00 & 1.00 & 2.34 & 6.00 & 2.00 & 1.41 & 220.74 & 578.26 & 99.92  \\ \hline
        20QBT\_45CYC\_0D1\_3D2\_9 & 20.00 & 135.00 & 1.00 & 2.33 & 6.00 & 2.00 & 1.41 & 180.00 & 420.29 & 99.69  \\ \hline
        20QBT\_45CYC\_1D1\_2D2\_6 & 20.00 & 180.00 & 0.50 & 2.54 & 6.00 & 1.00 & 1.00 & 63.89 & 183.61 & 98.19  \\ \hline
        20QBT\_45CYC\_1D1\_3D2\_7 & 20.00 & 225.00 & 0.60 & 2.34 & 6.00 & 2.00 & 1.41 & 105.78 & 252.63 & 99.88  \\ \hline
        20QBT\_45CYC\_2D1\_2D2\_3 & 20.00 & 270.00 & 0.33 & 2.55 & 6.00 & 1.00 & 1.00 & 11.48 & 109.16 & 97.89  \\ \hline
        20QBT\_45CYC\_3D1\_1D2\_8 & 20.00 & 315.00 & 0.14 & 2.88 & 5.00 & 1.00 & 0.00 & -50.79 & -2.20 & 78.94  \\ \hline
        20QBT\_45CYC\_4D1\_1D2\_3 & 20.00 & 405.00 & 0.11 & 2.95 & 5.00 & 1.00 & 0.00 & -59.26 & -31.86 & 80.48  \\ \hline
        3\_17\_13 & 3.00 & 36.00 & 0.47 & 1.00 & 2.00 & 2.00 & 2.83 & 111.11 & 341.03 & 21.96  \\ \hline
        4gt10-v1\_81 & 5.00 & 148.00 & 0.45 & 1.00 & 4.00 & 4.00 & 3.87 & 126.35 & 403.33 & 76.73  \\ \hline
        4gt11\_82 & 5.00 & 27.00 & 0.67 & 1.30 & 4.00 & 2.00 & 1.41 & 96.30 & 356.67 & 23.40  \\ \hline
        4gt13\_92 & 5.00 & 66.00 & 0.45 & 1.40 & 4.00 & 2.00 & 2.65 & 130.30 & 401.45 & 51.63  \\ \hline
        4gt5\_75 & 5.00 & 83.00 & 0.46 & 1.10 & 4.00 & 3.00 & 2.45 & 136.14 & 445.88 & 63.83  \\ \hline
        4mod5-bdd\_287 & 7.00 & 70.00 & 0.44 & 1.62 & 5.00 & 1.00 & 1.73 & 120.00 & 391.67 & 51.46  \\ \hline
        4mod7-v0\_94 & 5.00 & 162.00 & 0.44 & 1.10 & 4.00 & 3.00 & 4.47 & 126.54 & 419.39 & 80.25  \\ \hline
        adder\_n10 & 10.00 & 294.00 & 0.22 & 2.33 & 4.00 & 1.00 & 1.41 & 21.09 & 2027.27 & 85.25  \\ \hline
        adder\_n4 & 4.00 & 23.00 & 0.43 & 1.33 & 2.00 & 2.00 & 1.41 & 82.61 & 124.00 & 7.52  \\ \hline
        aj-e11\_165 & 5.00 & 151.00 & 0.46 & 1.00 & 4.00 & 4.00 & 4.36 & 123.84 & 407.24 & 76.73  \\ \hline
        alu-bdd\_288 & 7.00 & 84.00 & 0.45 & 1.43 & 5.00 & 2.00 & 1.73 & 125.00 & 356.47 & 61.65  \\ \hline
        alu-v0\_26 & 5.00 & 84.00 & 0.45 & 1.20 & 4.00 & 3.00 & 3.16 & 115.48 & 376.47 & 53.61  \\ \hline
        alu-v1\_28 & 5.00 & 37.00 & 0.49 & 1.30 & 4.00 & 2.00 & 1.41 & 108.11 & 342.11 & 25.31  \\ \hline
        alu-v3\_34 & 5.00 & 52.00 & 0.46 & 1.30 & 4.00 & 2.00 & 1.73 & 125.00 & 394.55 & 41.42  \\ \hline
        alu-v4\_36 & 5.00 & 115.00 & 0.44 & 1.10 & 4.00 & 3.00 & 3.32 & 117.39 & 358.47 & 62.04  \\ \hline
        basis\_change\_n3 & 3.00 & 79.00 & 0.13 & 1.33 & 2.00 & 1.00 & 2.65 & 0.00 & 46.91 & 0.00  \\ \hline
        bell\_n4 & 4.00 & 49.00 & 0.14 & 1.67 & 2.00 & 1.00 & 1.00 & 8.16 & 49.02 & 2.88  \\ \hline
        bv\_n14 & 14.00 & 41.00 & 0.32 & 1.86 & 13.00 & 1.00 & 0.00 & 146.34 & 280.95 & 58.41  \\ \hline
        bv\_n19 & 19.00 & 56.00 & 0.32 & 1.89 & 18.00 & 1.00 & 0.00 & 178.57 & 349.12 & 79.99  \\ \hline
        cat\_state\_n4 & 4.00 & 4.00 & 0.75 & 1.67 & 2.00 & 1.00 & 0.00 & 25.00 & 100.00 & 0.18  \\ \hline
        cuccaroAdder\_1b & 4.00 & 73.00 & 0.23 & 1.33 & 3.00 & 1.00 & 1.41 & 6.85 & 1540.00 & 20.45  \\ \hline
        cuccaroMultiplier\_1b & 6.00 & 176.00 & 0.18 & 1.20 & 4.00 & 2.00 & 1.00 & 8.52 & 944.12 & 47.40  \\ \hline
        decod24-v1\_41 & 5.00 & 85.00 & 0.45 & 1.20 & 4.00 & 2.00 & 2.65 & 124.71 & 409.09 & 57.27  \\ \hline
        deutsch\_n2 & 2.00 & 5.00 & 0.20 & 1.00 & 1.00 & 1.00 & 0.00 & 20.00 & 14.29 & 0.18  \\ \hline
        dnn\_n2 & 2.00 & 338.00 & 0.12 & 1.00 & 1.00 & 1.00 & 24.25 & 0.00 & 97.65 & 0.00  \\ \hline
        error\_correctiond3\_n5 & 5.00 & 114.00 & 0.43 & 1.50 & 4.00 & 1.00 & 5.92 & 60.53 & 139.13 & 24.28  \\ \hline
        ex-1\_166 & 3.00 & 19.00 & 0.47 & 1.00 & 2.00 & 2.00 & 1.41 & 105.26 & 319.05 & 11.66  \\ \hline
        fredkin\_n3 & 3.00 & 19.00 & 0.42 & 1.00 & 2.00 & 2.00 & 1.41 & 94.74 & 209.09 & 7.35  \\ \hline
        graycode6\_47 & 6.00 & 5.00 & 1.00 & 2.33 & 2.00 & 1.00 & 0.00 & 80.00 & 375.00 & 9.07  \\ \hline
        grover\_n2 & 2.00 & 16.00 & 0.13 & 1.00 & 1.00 & 1.00 & 1.00 & -43.75 & -23.53 & -1.27  \\ \hline
        grover\_n3 & 3.00 & 89.00 & 0.13 & 1.00 & 2.00 & 2.00 & 1.00 & -19.10 & 945.45 & 11.59  \\ \hline
        iswap\_n2 & 2.00 & 9.00 & 0.22 & 1.00 & 1.00 & 1.00 & 1.00 & 44.44 & 27.27 & 0.72  \\ \hline
        linearsolver\_n3 & 3.00 & 23.00 & 0.17 & 1.33 & 2.00 & 1.00 & 1.00 & 8.70 & 56.00 & 0.36  \\ \hline
        lpn\_n5 & 5.00 & 11.00 & 0.18 & 1.33 & 2.00 & 1.00 & 0.00 & 18.18 & 50.00 & 2.53  \\ \hline
        miller\_11 & 3.00 & 50.00 & 0.46 & 1.00 & 2.00 & 2.00 & 3.87 & 120.00 & 373.58 & 31.05  \\ \hline
        mod5d2\_64 & 5.00 & 53.00 & 0.47 & 1.20 & 4.00 & 2.00 & 1.73 & 115.09 & 420.00 & 43.53  \\ \hline
        multiply\_n13 & 11.00 & 140.00 & 0.19 & 2.44 & 4.00 & 1.00 & 0.00 & 24.29 & 1854.55 & 63.44  \\ \hline
        one-two-three-v1\_99 & 5.00 & 132.00 & 0.45 & 1.10 & 4.00 & 3.00 & 3.61 & 115.15 & 371.11 & 67.01  \\ \hline
        q=10\_s=50\_2qbf=096\_1 & 10.00 & 60.00 & 0.80 & 1.36 & 8.00 & 3.00 & 1.00 & 76.67 & 242.62 & 66.51  \\ \hline
        q=10\_s=90\_2qbf=011\_1 & 10.00 & 100.00 & 0.10 & 2.67 & 3.00 & 1.00 & 0.00 & -1.00 & 14.00 & 47.04  \\ \hline
        q=11\_s=49\_2qbf=061\_1 & 11.00 & 60.00 & 0.52 & 1.62 & 7.00 & 2.00 & 0.00 & 86.67 & 234.43 & 72.21  \\ \hline
        q=11\_s=89\_2qbf=022\_1 & 11.00 & 100.00 & 0.17 & 1.84 & 6.00 & 1.00 & 0.00 & 39.00 & 68.32 & 67.54  \\ \hline
        q=17\_s=43\_2qbf=028\_1 & 17.00 & 60.00 & 0.20 & 3.97 & 3.00 & 1.00 & 0.00 & 53.33 & 90.00 & 55.29  \\ \hline
        q=3\_s=57\_2qbf=011\_1 & 3.00 & 60.00 & 0.05 & 1.33 & 2.00 & 1.00 & 0.00 & -6.67 & -58.33 & -0.72  \\ \hline
        q=3\_s=97\_2qbf=01\_1 & 3.00 & 100.00 & 0.10 & 1.00 & 2.00 & 2.00 & 1.41 & -10.00 & -5.00 & 4.68  \\ \hline
        q=4\_s=56\_2qbf=032\_1 & 4.00 & 60.00 & 0.28 & 1.00 & 3.00 & 3.00 & 1.00 & 3.33 & 65.00 & 14.57  \\ \hline
        q=4\_s=96\_2qbf=052\_1 & 4.00 & 100.00 & 0.54 & 1.00 & 3.00 & 3.00 & 5.20 & 12.00 & 133.66 & 31.15  \\ \hline
        q=5\_s=55\_2qbf=087\_1 & 5.00 & 60.00 & 0.85 & 1.00 & 4.00 & 4.00 & 3.16 & 15.00 & 139.34 & 21.02  \\ \hline
        q=5\_s=95\_2qbf=095\_1 & 5.00 & 100.00 & 0.90 & 1.00 & 4.00 & 4.00 & 4.36 & 13.00 & 114.00 & 26.59  \\ \hline
        q=6\_s=54\_2qbf=022\_1 & 6.00 & 60.00 & 0.23 & 1.47 & 3.00 & 2.00 & 0.00 & 26.67 & 150.00 & 37.40  \\ \hline
        q=6\_s=94\_2qbf=053\_1 & 6.00 & 100.00 & 0.52 & 1.00 & 5.00 & 5.00 & 2.24 & 31.00 & 221.78 & 58.07  \\ \hline
        q=7\_s=53\_2qbf=034\_1 & 7.00 & 60.00 & 0.33 & 1.33 & 6.00 & 2.00 & 0.00 & 23.33 & 101.67 & 41.18  \\ \hline
        q=7\_s=93\_2qbf=054\_1 & 7.00 & 100.00 & 0.52 & 1.14 & 6.00 & 4.00 & 1.73 & 78.00 & 244.55 & 84.35  \\ \hline
        q=8\_s=52\_2qbf=104\_1 & 8.00 & 60.00 & 0.87 & 1.11 & 7.00 & 5.00 & 1.00 & 51.67 & 213.11 & 52.15  \\ \hline
        q=8\_s=92\_2qbf=011\_1 & 8.00 & 100.00 & 0.17 & 1.57 & 5.00 & 1.00 & 0.00 & 20.00 & 50.00 & 58.16  \\ \hline
        q=9\_s=51\_2qbf=012\_1 & 9.00 & 60.00 & 0.12 & 2.29 & 3.00 & 1.00 & 0.00 & 10.00 & 25.00 & 28.86  \\ \hline
        q=9\_s=51\_2qbf=059\_1 & 9.00 & 60.00 & 0.47 & 1.47 & 6.00 & 2.00 & 0.00 & 68.33 & 170.00 & 63.10  \\ \hline
        q=9\_s=91\_2qbf=088\_1 & 9.00 & 100.00 & 0.81 & 1.08 & 8.00 & 5.00 & 1.41 & 69.00 & 248.51 & 80.62  \\ \hline
        qaoa\_n16 & 16.00 & 372.00 & 0.52 & 1.20 & 12.00 & 12.00 & 0.00 & 90.59 & 192.23 & 99.91  \\ \hline
        qaoa\_n6 & 6.00 & 414.00 & 0.13 & 1.40 & 3.00 & 3.00 & 3.00 & 6.52 & 75.48 & 39.96  \\ \hline
        qec\_en\_n5 & 5.00 & 25.00 & 0.40 & 1.60 & 4.00 & 1.00 & 1.00 & 64.00 & 167.86 & 11.02  \\ \hline
        qec\_sm\_n5 & 5.00 & 25.00 & 0.40 & 1.60 & 4.00 & 1.00 & 1.00 & 64.00 & 167.86 & 11.02  \\ \hline
        qft\_n15 & 15.00 & 540.00 & 0.39 & 1.00 & 14.00 & 14.00 & 0.00 & 30.37 & 112.75 & 97.19  \\ \hline
        qft\_n20 & 20.00 & 970.00 & 0.39 & 1.00 & 19.00 & 19.00 & 0.00 & 42.16 & 154.79 & 99.99  \\ \hline
        quantum\_volume\_n5 & 5.00 & 338.00 & 0.15 & 1.20 & 4.00 & 2.00 & 6.93 & 6.80 & 91.18 & 40.84  \\ \hline
        rd32\_270 & 5.00 & 84.00 & 0.43 & 1.10 & 4.00 & 3.00 & 2.45 & 117.86 & 408.05 & 57.67  \\ \hline
        seca\_n11 & 11.00 & 333.00 & 0.25 & 1.95 & 6.00 & 2.00 & 1.73 & 11.71 & 590.00 & 81.67  \\ \hline
        shor\_15 & 11.00 & 4792.00 & 0.37 & 1.38 & 8.00 & 4.00 & 36.61 & 130.74 & 267.26 & 100.00  \\ \hline
        shor\_35 & 15.00 & 16529.00 & 0.37 & 1.41 & 12.00 & 5.00 & 79.04 & 130.82 & 256.91 & 100.00  \\ \hline
        simon\_n6 & 5.00 & 82.00 & 0.17 & 1.70 & 3.00 & 1.00 & 0.00 & -9.76 & 635.29 & 14.91  \\ \hline
        teleportation\_n3 & 3.00 & 8.00 & 0.25 & 1.33 & 2.00 & 1.00 & 0.00 & 62.50 & 50.00 & 0.90  \\ \hline
        toffoli\_n3 & 3.00 & 18.00 & 0.33 & 1.00 & 2.00 & 2.00 & 1.00 & 116.67 & 245.00 & 9.86  \\ \hline
        variational\_n4 & 4.00 & 54.00 & 0.30 & 1.67 & 2.00 & 1.00 & 2.83 & 14.81 & 103.57 & 1.43  \\ \hline
        vbeAdder\_1b & 4.00 & 70.00 & 0.20 & 1.17 & 3.00 & 2.00 & 1.00 & 5.71 & 1928.57 & 20.31  \\ \hline
        vqe\_uccsd\_n4 & 4.00 & 220.00 & 0.40 & 1.67 & 2.00 & 1.00 & 14.70 & -12.73 & 110.86 & -5.17  \\ \hline
        wstate\_n3 & 3.00 & 49.00 & 0.18 & 1.00 & 2.00 & 2.00 & 1.41 & 12.24 & 363.16 & 9.40  \\ \hline
        xor5\_254 & 6.00 & 7.00 & 0.71 & 1.67 & 5.00 & 1.00 & 0.00 & 85.71 & 400.00 & 13.30 \\ \hline
    \end{tabularx}
    \label{tab:results_ibmrochester}
\end{table}

\end{document}